\documentclass{article}

\usepackage{arxiv}

\usepackage[utf8]{inputenc}       
\usepackage[T1]{fontenc}          
\usepackage{hyperref}             
\usepackage{url}                  
\usepackage{booktabs}             
\usepackage{amsmath}
\usepackage{amsfonts}             
\usepackage{amssymb}
\usepackage{nicefrac}             
\usepackage{microtype}            
\usepackage{graphicx}
\usepackage[table]{xcolor}
\usepackage{multirow}
\usepackage[authoryear,longnamesfirst]{natbib}
\usepackage{doi}

\definecolor{staisblue}{HTML}{DBEAFE}
\definecolor{ntamber}{HTML}{FEF3C7}
\definecolor{retgrey}{HTML}{F3F4F6}

\title{Generative AI Steering and Human Authoring: Complementary
  Search-as-Learning Scaffolds in Cross-Session Aggregated Academic Search}

\author{
  \href{https://orcid.org/0000-0003-4267-4938}{\includegraphics[scale=0.06]{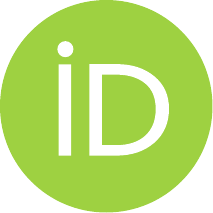}\hspace{1mm}Milad
    Momeni}\thanks{Corresponding author.} \\
  Department of Computer Science \\
  University of Regina \\
  \texttt{miladmomeni@uregina.ca} \\
  \And
  \href{https://orcid.org/0000-0003-2875-1132}{\includegraphics[scale=0.06]{orcid.pdf}\hspace{1mm}Orland
    Hoeber} \\
  Department of Computer Science \\
  University of Regina \\
  \texttt{orland.hoeber@uregina.ca} \\
}

\date{}

\makeatletter
\renewcommand{\@maketitle}{%
  \vbox{%
    \hsize\textwidth
    \linewidth\hsize
    \vskip 0.1in
    \@toptitlebar
    \centering
    {\LARGE\bfseries \@title\par}
    \@bottomtitlebar
    \undertitle\\
    \vskip 0.1in
    \def\And{%
      \end{tabular}\hfil\linebreak[0]\hfil%
      \begin{tabular}[t]{c}\bf\rule{\z@}{24\p@}\ignorespaces%
    }
    \def\AND{%
      \end{tabular}\hfil\linebreak[4]\hfil%
      \begin{tabular}[t]{c}\bf\rule{\z@}{24\p@}\ignorespaces%
    }
    \begin{tabular}[t]{c}\bf\rule{\z@}{24\p@}\@author\end{tabular}%
    \vskip 0.4in \@minus 0.1in \center{\@date} \vskip 0.2in
  }
}
\makeatother

\renewcommand{\headeright}{Preprint}
\renewcommand{\undertitle}{Preprint: accepted at the Information Processing \& Management Conference 2026 (IP\&MC2026)}
\renewcommand{\shorttitle}{GenAI Steering and Human Authoring in Cross-Session Academic Search}

\hypersetup{
  colorlinks=true,
  linkcolor=[rgb]{0.10,0.25,0.55},
  citecolor=[rgb]{0.10,0.25,0.55},
  urlcolor=[rgb]{0.10,0.25,0.55},
  pdftitle={Generative AI Steering and Human Authoring: Complementary Search-as-Learning Scaffolds in Cross-Session Aggregated Academic Search},
  pdfauthor={Milad Momeni, Orland Hoeber},
  pdfkeywords={GenAI summarization, Cross-session search, Search task resumption, Reacquaintance, Search-as-learning, Aggregated search, Academic search, Exploratory search},
}

\begin{document}
\maketitle

\begin{abstract}
Aggregated academic search interfaces combine resources from heterogeneous sources, such as scholarly libraries and reference encyclopedias, into a single workspace. Such tasks frequently span multiple sessions, requiring learners to return to and resume work after hours or days. End-of-session \emph{consolidation}, the activity of organizing and synthesizing gathered information so it can be resumed later, is important in this setting but underexplored in search-as-learning. Generative AI (GenAI) has the potential to make consolidation more efficient, but it may also shift cognitive work away from the learner. This study compares two consolidation scaffolds embedded in the workspace framework of an existing academic search interface that aggregates and clusters saved search results from two different sources (our university library and Wikipedia), which was developed in earlier work.

\textit{STAIS} (Steerable AI Summaries) provides a workspace enhancement that produces GenAI summaries that users can correct, reorganize, and fact-check, automating parts of the consolidation activity. \textit{NT} (Note-Taking) provides a workspace enhancement that includes rich-text note editors that learners can use to author their own consolidations of what they have found. In a controlled between-subjects study ($n=36$), participants completed two sessions one week apart while searching an unfamiliar topic using either \textit{STAIS} or \textit{NT}. The two conditions produced contrasting post-resumption outcome profiles: \textit{STAIS} supported faster task resumption and was associated with broader conceptual coverage and higher self-reported domain knowledge, whereas \textit{NT} was associated with richer inter-concept connections and higher self-reported certainty. Rather than uniformly helping or harming learning, the two scaffolds shaped which facets of post-resumption learning were more strongly supported, positioning the division of consolidation labour between system and learner as a key design variable for GenAI-supported search-as-learning scaffolds.
\end{abstract}


\renewcommand{\thefootnote}{\arabic{footnote}}
\setcounter{footnote}{0}

\section{Introduction}
\label{sec:intro}

Academic search rarely unfolds in a single session~\citep{belkin1995cases,li2020everyday,morris2008searchbar,agichtein2012search,li2024stopped}. Students preparing a research brief, practitioners surveying an unfamiliar domain, and researchers entering a new field often distribute their information work across multiple sessions separated by hours or days. They suspend or stop those sessions for many reasons beyond external interruption alone~\citep{li2024stopped}. When they return, they must reconstruct enough context to continue productively: identify what they had already found, understand how they had organized it, recall what seemed important, and assess what questions remained open~\citep{altmann2002memory,trafton2003preparing}. The external cues available at that moment, such as saved resources, workspace structure, notes, and reminders, strongly shape how quickly and how well the task can be resumed~\citep{capra2013augmenting,li2024stopped}. In aggregated academic search settings, where evidence is drawn from heterogeneous sources such as scholarly libraries and reference encyclopedias~\citep{arguello2017aggregated} and merged into a single workspace~\citep{gomes2022study,orin2025organizing}, this reacquaintance challenge is amplified by the volume and diversity of saved material that users must reintegrate.

In exploratory search, resumption is not only a memory problem. Returning to a task also requires reconnecting with what has already been learned and reorganizing it into a form that supports further inquiry. This makes \emph{consolidation} (how encountered information is selected, organized, and externalized into reusable representations) a shared concern of two research areas that are often studied separately: search-as-learning (SAL), which examines how search changes what a learner knows~\citep{marchionini2006exploratory,urgo2025search}, and cross-session search, which examines how systems support return to interrupted or suspended tasks~\citep{capra2010tools-at-hand,crescenzi2021supporting,li2024stopped}. 

Throughout this paper, \emph{consolidation} refers specifically to externalizing and organizing activities, which learners perform near the end of a search session in preparation for resuming it later. This is distinct from any offline cognitive memory consolidation that may occur during the inter-session interval, which is not measured here. From a learning perspective, such tasks often require conceptual understanding and analysis; consolidation choices shape both reacquaintance and post-resumption learning. 
Yet most search interfaces offer only limited support for consolidation activities. Searchers can save, label, and sometimes annotate resources, but the burden of synthesizing, restructuring, and preserving what has been learned remains largely a manual effort.

Generative AI (GenAI) opens many opportunities within this design space. Large language models can summarize saved resources and provide compact cues that support task resumption~\citep{orin2025organizing}. Prior work has shown that GenAI cluster summaries can improve resumption efficiency within an aggregated digital humanities search context~\citep{momeni2025crosssession}.
What remains unclear is how delegating part of consolidation to the system affects learning after resumption. This question is especially important in educational and research-oriented search, where faster reacquaintance is valuable, but not if it comes at the cost of weaker knowledge construction. Related studies of GenAI-assisted learning have largely focused on single-session settings or comparing GenAI answers with conventional search~\citep{bastani2025generative,melumad2025llm,carmi2025genailearning}, rather than examining how different consolidation scaffolds shape learning across sessions.

Whether this distinction matters depends on how consolidation effort relates to learning. If durable knowledge requires active construction and effortful retrieval~\citep{tulving1962subjective,bransford2000people,roediger2006test,karpicke2011retrieval}, then delegating consolidation to GenAI may weaken encoding. Yet GenAI support need not be purely passive: when learners inspect, correct, and verify GenAI summaries, consolidation effort may be redistributed rather than removed. The central question is therefore not simply whether GenAI helps searchers resume their previously suspended search tasks faster, but whether steerable GenAI-supported consolidation yields a different \emph{post-resumption learning} profile 
from human-authored note-taking, and which facets of learning each scaffold most strongly supports.

To examine this question, we compare two conditions within an existing aggregated academic search interface that combines results from our university's academic library and Wikipedia. The interface includes a multi-workspace framework for academic search~\citep{gomes2022study}; the aggregated-search and clustering layer used here was developed and studied in our prior work~\citep{momeni2025study}. Both conditions provide the same search interface, clustering, organizational controls, and persistence across sessions. They differ only in the nature of the consolidation scaffold. In \textit{NT} (\textit{Note-Taking}), users create their own cluster notes, drawing on evidence that human-authored note-taking supports both encoding and later retrieval in search-as-learning contexts~\citep{urgo2025search,capra2010tools-at-hand,roy2021notehighlight}. In \textit{STAIS} (\textit{Steerable AI Summaries}), the system generates cluster-level summaries that users can steer through inline correction, structure-driven regeneration, and claim-level fact-checking, thereby preserving a degree of learner agency during consolidation~\citep{heer2019agency,amershi2019guidelines}. We evaluate these alternatives in a controlled between-subjects study ($n=36$) conducted across two sessions separated by one week. Participants searched an unfamiliar topic, consolidated what they found at the end of the initial session, and returned in the resumed session to continue the same search task; they finished by writing a short synthesis of the topic. Because this assessment was administered at the end of the resumed session, the outcomes reflect post-resumption learning shaped by both prior consolidation and subsequent search, rather than delayed retention of the initial session alone.

This paper makes three contributions. First, at a \emph{conceptual} level, it frames the allocation of consolidation labour between learner and system as a first-class design variable for GenAI-supported search-as-learning scaffolds, connecting SAL, cross-session search, and human-centred AI perspectives that have previously been treated in isolation~\citep{carmi2025genailearning}. Second, at an \emph{artifact} level, it introduces \textit{STAIS}, a workspace that combines GenAI cluster summaries with user steering through inline correction, structure-driven regeneration, and claim-level fact-checking, operationalizing the consolidation-labour dimension within a functional cross-session aggregated academic search environment. Third, at an \emph{empirical} level, it reports results from a controlled two-session study ($n=36$) showing that human-authored and GenAI-steered consolidation produce complementary post-resumption outcome profiles: \textit{STAIS} was associated with broader conceptual coverage and faster resumption, whereas \textit{NT} was associated with richer inter-concept connections and higher self-reported certainty.
Rather than asking whether GenAI simply helps or harms learning, this work examines how different allocations of consolidation work between system and learner shape which facets of post-resumption learning are most strongly supported.

\section{Related Work}
\label{sec:related}

This study draws on three intersecting areas of research: search-as-learning (SAL), cross-session search, and GenAI-assisted sensemaking. Across the literature, a common issue is consolidation, which can shape both how effectively users resume a suspended task and what they retain from it over time. Prior work has established the value of persistent workspace context, and more recently, GenAI-generated summaries for cross-session search. However, it remains unclear how different allocations of consolidation work between learner and system affect post-resumption learning. 

\subsection{Search-as-Learning and the Anderson and Krathwohl (A\&K) Taxonomy}

Search-as-learning (SAL) research treats information seeking not only as retrieval, but also as a process through which people build knowledge and understanding in the course of searching~\citep{kuhlthau1991search,marchionini2006exploratory,crescenzi2021supporting,urgo2025search}. Within this tradition, the revised taxonomy of \citep{anderson2001taxonomy} (A\&K) provides a useful lens for both task design and outcome interpretation because it distinguishes among knowledge types (from factual to metacognitive) and cognitive processes (from remember to create)~\citep{urgo2019assessment,urgo2025search,srinivas2025examining}. For open-ended academic search tasks, the most relevant targets often lie in \emph{conceptual knowledge} and in the processes of \emph{understand} and \emph{analyze}, where learners must identify key ideas, reconcile evidence across sources, compare perspectives, and explain relationships in their own words. We refer to specific knowledge $\times$ cognitive-process intersections as A\&K \emph{cells} (e.g., \textit{Understand} $\times$ \textit{Conceptual}), following the terminology used in SAL research~\citep{urgo2025search}.

These demands make consolidation especially important. In SAL settings, a simple record of isolated facts may support later review, but deeper conceptual learning often depends on activities that require learners to connect, reorganize, and interpret ideas across sources. This distinction is particularly relevant because it has been identified that delayed retention and knowledge transfer are outcomes that remain underexplored in SAL research~\citep{urgo2025search}. Related work in learning science also suggests that effortful processing can support stronger post-search retention and knowledge transfer than passive review. Retrieval practice has been shown to strengthen long-term retention~\citep{roediger2006test,karpicke2011retrieval,roediger2011retrieval}, and the desirable difficulties perspective argues that appropriately effortful processing can improve later memory and understanding~\citep{bjork1994desirable}. Taken together, these arguments suggest that consolidation scaffolds that require learners to generate, select, and organize ideas may support post-resumption learning differently from scaffolds that primarily facilitate passive review.

Open-ended written responses are well-suited to capturing such differences because they can reveal both breadth of conceptual coverage and depth of conceptual integration~\citep{wilson2013assessment,roy2021notehighlight}. A two-session design further allows learning to be examined across time rather than within a single session. Once learning is studied across sessions, however, a practical question arises: when learners return to a suspended task after a delay, how can a workspace support reacquaintance, and how do the representations created earlier shape subsequent learning? 

\subsection{Cross-Session Search, Workspace Organization, and Learner-Authored Consolidation}

Complex information tasks often unfold across multiple sessions rather than within a single continuous interaction~\citep{white2009exploratory}. Studies of everyday cross-session search show that users routinely distribute information work across sessions separated by hours or days, and that the reasons for resumption range from acquiring new information needs to revisiting previously incomplete tasks~\citep{li2020everyday,agichtein2012search}. When users return, they must often reconstruct what they examined, why resources were saved, how information had been organized, and what remains to be done~\citep{capra2010tools-at-hand}. The quality of available context cues strongly affects reacquaintance and continuation: memory-for-goals theory predicts that the strength and accessibility of contextual cues determine how quickly a suspended goal can be retrieved~\citep{altmann2002memory,trafton2003preparing}, and recent diary studies confirm that richer cues reduce the perceived difficulty of returning to tasks~\citep{li2024stopped}. Working memory also plays a role; \citet{choi2025effects} found that individual differences in working memory capacity influence how effectively users manage information during search and sensemaking, suggesting that the cognitive demands of resumption are not uniform across users. For this reason, prior systems have aimed to reduce resumption costs by preserving search histories, saved artifacts, and other forms of workspace context~\citep{morris2008searchbar,xu2018logcanvas,crescenzi2021supporting,gomes2022study,orin2025organizing,hoeber-islam-2026}.

A recurring observation in this literature is that the \emph{type} of context cue matters. Query logs and browser histories provide chronological traces but offer limited semantic structure; saved-resource collections preserve content but not the user's interpretation of it; and annotated workspaces provide richer cues but require upfront effort to create. What has received less attention is whether cues \emph{authored by the learner} differ in their effects from cues \emph{generated by the system}, even when both are available in the same workspace at resumption time~\citep{orin2025organizing}. That distinction may matter because human-authored representations can serve two roles simultaneously, functioning not only as external cues for later resumption but also as encoding activities at the time of creation. In aggregated search settings~\citep{arguello2017aggregated,gomes2022study}, where users collect material from multiple sources and impose structure on it, this issue becomes especially salient. Cluster-based workspaces help users manage and revisit saved materials~\citep{cutting1992scattergather}, but visible structure alone may not fully explain later retrieval. Tulving's work on subjective organization suggests that the act of imposing one's own structure on information can itself support later recall~\citep{tulving1962subjective}. A representation constructed by the learner may therefore function differently from one generated automatically, even when both appear similarly useful at the interface level.

This distinction is particularly relevant for note-taking and annotation. Human-authored notes are among the most established consolidation mechanisms in SAL and related learning research~\citep{marshall1997annotation,urgo2025search}. Their value lies not only in producing a record, but also in requiring cognitive processing through selecting, paraphrasing, organizing, and linking ideas~\citep{mueller2014pen}. More broadly, writing-to-learn research shows that analytic writing tasks, those requiring inference and connection-making rather than simple recording, tend to improve learning outcomes across domains~\citep{bangertdrowns2004writing}. Within search contexts, note-taking tools have also been associated with improved downstream fact coverage in writing tasks~\citep{roy2021notehighlight}. At the same time, note-taking is effortful and time-consuming, and not all learners benefit equally from having to construct useful external representations on their own~\citep{urgo2026sal}. This creates an important design tension for cross-session learning environments: human-authored consolidation may strengthen encoding, but it also imposes sustained effort that some learners may struggle to use effectively~\citep{bastani2025generative}.

Recent work on cross-session aggregated search in digital humanities contexts has shown that GenAI-generated cluster summaries can improve task resumption efficiency in organized workspaces~\citep{momeni2025study,momeni2025crosssession}.
These results establish that summaries can help users return to task, though they may not automatically generalize to other academic search settings. They do not yet show whether human-authored and GenAI-assisted consolidation differ in their effects on post-resumption conceptual learning.

\subsection{Generative AI for Consolidation, Sensemaking, and User Agency}

GenAI offers a new way to support consolidation by producing summaries, thematic overviews, and other synthesized representations of a user's saved materials~\citep{liu2024selenite}. In the broader sensemaking literature, consolidation corresponds to the stages in which users move from collected evidence toward structured interpretation and explanation~\citep{pirolli2005sensemaking}. Retrieval-augmented generation (RAG) can ground generated output in source evidence~\citep{lewis2020retrieval-augmented}, and workspace-augmented generation (WAG) extends this idea by generating summaries from materials that searchers have put the effort into saving and organizing during the task~\citep{orin2025organizing}. From a systems perspective, this is appealing because it can reduce the burden of rereading, relocating, and mentally reintegrating previously encountered information, thereby providing compact cues that may support faster resumption.

From a learning perspective, however, the implications are less straightforward. The ICAP (Interactive--Constructive--Active--Passive) framework~\citep{chi2014icap} distinguishes four levels of cognitive engagement.
The four levels are \emph{passive} (receiving information without overt processing), \emph{active} (manipulating or selecting within provided material), \emph{constructive} (generating novel output that goes beyond the given information), and \emph{interactive} (co-constructing meaning through dialogue or joint production). ICAP predicts that learning outcomes improve as engagement moves from passive toward constructive or interactive modes. Writing one's own notes is a prototypically \emph{constructive} activity because the learner generates novel output; reading a system-generated summary without modification is largely \emph{passive}; and selectively editing, correcting, or restructuring GenAI-generated content may place engagement at an \emph{active} to \emph{constructive} level depending on the depth of the learner's involvement. 

Adjacent empirical work suggests that this concern deserves careful examination. In some settings, LLM-generated answers have been associated with shallower knowledge than conventional web search when users invest less reasoning effort~\citep{melumad2025llm}. GenAI-assisted study has in some cases produced weaker delayed outcomes than an unassisted study~\citep{bastani2025generative}. GenAI-generated summaries also carry a well-documented risk of factual distortion, with confabulation (commonly misrepresented as hallucination) remaining a persistent challenge even in current frontier models~\citep{maynez2020faithfulness,huang2024hallucination}. These results do not come from the same cross-session search context studied here, but they caution against assuming that efficiency gains automatically translate into stronger learning.

At the same time, reduced effort is not inherently undesirable. Cognitive Load Theory (CLT) distinguishes between extraneous load, which arises from avoidable cognitive processing that does not contribute to learning, and germane load, which reflects effort devoted to schema construction and integration~\citep{sweller2010element}. From this perspective, a scaffold may support learning if it reduces extraneous effort while preserving the learner's responsibility to evaluate, connect, and interpret ideas, which is precisely the kind of effortful, generative processing that the desirable difficulties perspective treats as building durable understanding~\citep{bjork1994desirable}. In the present context, GenAI summaries may help by reducing the overhead of reacquainting users with previously saved materials, while still leaving room for deeper engagement through inspection, comparison, correction, and revision. However, CLT also predicts that the benefit of scaffolding depends on the learner's existing knowledge: the expertise reversal effect shows that instructional supports helpful for novices can become redundant or even counterproductive for more knowledgeable learners~\citep{kalyuga2003expertise}. Recent SAL studies that focus on novice searchers learning unfamiliar scientific topics~\citep{urgo2026sal,ghosh2018searching} suggest that scaffolding benefits are most likely when prior knowledge is low, but the expertise reversal effect highlights that GenAI-assisted consolidation cannot be assumed to benefit all learners equally across contexts.

Whether this balance is achieved depends heavily on user agency. Human-centred AI research consistently argues for systems that preserve meaningful user control, inspection, and correction~\citep{heer2019agency,amershi2019guidelines,shneiderman2020human-centered}. Related work also shows that lightweight steering mechanisms can improve generated output while keeping users actively involved in the process~\citep{tang2024steering,tang2025ReSPIRE,wu2021recursively}. This makes user agency central to GenAI-assisted consolidation. Passive reading of a summary is likely to remain relatively shallow, whereas evaluating, editing, correcting, or restructuring GenAI-generated content may, as noted earlier, move engagement toward the active or constructive end of the ICAP continuum~\citep{chi2014icap}. What remains unclear is whether this reactive form of engagement yields learning benefits comparable to, or different from, the more proactive generative processing involved in authorship. Existing GenAI-assisted sensemaking systems primarily emphasize output quality, controllability, and interaction design~\citep{liu2024selenite,tang2025ReSPIRE}; SAL studies, by contrast, focus on learning outcomes but rarely compare human-authored and GenAI-assisted consolidation within the same cross-session workspace.

\subsection{Synthesis: Prior Work and Research Questions}

Taken together, this literature review points to a central unresolved tension. ICAP implies that offloading cognitive work to the system may reduce engagement below the constructive threshold, whereas CLT suggests that removing extraneous reacquaintance effort may free cognitive resources for deeper integration. Resolving this tension empirically requires a design that holds the search environment constant while varying only the consolidation scaffold, and that measures learning outcomes after resumption rather than within a single session. Prior studies have not met both requirements simultaneously: as noted earlier, GenAI-assisted sensemaking systems evaluate output quality and interaction but not post-resumption learning~\citep{liu2024selenite,tang2025ReSPIRE}; SAL studies measure learning but typically within single sessions and without comparing consolidation origins~\citep{urgo2025search,urgo2026sal,ghosh2018searching}; and cross-session search studies evaluate resumption efficiency but not the learning consequences~\citep{gomes2022study,hoeber-islam-2026}.

The present study addresses this gap by embedding both a human-authored note-taking condition (\textit{NT}) and a steerable GenAI-summary condition (\textit{STAIS}) within the same aggregated academic search interface. This design isolates the consolidation scaffold while keeping the foundational search/workspace environment constant.

\section{System Design}
\label{sec:design}

\subsection{Shared Design Foundation}

The design of the two interface alternatives for scaffolding SAL was guided by two goals, both of which were informed by design principles for exploratory search interfaces~\citep{hoeber2025design}: (1) supporting continuity across sessions by helping users preserve, revisit, and reorganize what they had already found, and (2) supporting active engagement with saved materials rather than encouraging passive accumulation or passive consumption of GenAI output~\citep{carmi2025genailearning}.

Both conditions build on the same aggregated academic search framework for cross-session searching. The underlying multi-workspace framework was introduced by~\citet{gomes2022study}; the aggregated-search and clustering layer used here, which integrates an academic library with a reference encyclopedia, was added in our prior work~\citep{momeni2025study,momeni2025crosssession}. Aggregated search integrates results from multiple heterogeneous sources into a unified interface, removing the need to query each source separately~\citep{arguello2017aggregated}. In the present context, the system aggregates results from two complementary platforms: our university's academic library and Wikipedia~\citep{librisgroup2025primo,mediawiki2025api}. The academic library provides peer-reviewed and scholarly resources, while Wikipedia supplies accessible, fact-oriented background that can help learners build an initial understanding of unfamiliar topics. Prior research has shown that Wikipedia articles are broadly reliable for factual coverage and are widely used as entry points in academic information seeking~\citep{mesgari2015sum,singer2017whyweread}. Presenting both source types on a single interface reduces the platform-switching overhead that can fragment exploratory search~\citep{capra2010tools-at-hand,yang2025searchchat}. The system intentionally excludes conversational GenAI (e.g., ChatGPT) as a retrieval source, because the study focuses on how learners consolidate \emph{found} information from verifiable and citable documents rather than on how they interact with generated text whose provenance cannot be independently traced. For each query, the system retrieves results in parallel from both sources, presents ten results per source per page of results, and merges them in round-robin order on a unified search engine results page (SERP)~\citep{chuklin2015interleaving}. Each result card displays its source icon and includes a \emph{save} button so that potentially useful materials can be preserved in the workspace.

When users save resources, the workspace organizes them into thematic clusters. The cluster structure is generated incrementally rather than in a single pass: the first saved resource forms an initial cluster, and each subsequently saved resource is assigned by a large language model (GPT-4o-mini;~\citealt{openai2024gpt4omini}) to the existing cluster whose accumulated content best matches it, or, when no existing cluster fits, used to seed a new cluster (see Appendix~\ref{app:implementation} for full technical details). After each assignment, the model generates a short keyword-style label of up to four words for the affected cluster. This automated organization provides structure for the workspace, while preserving continuity as new materials are added. In addition, the workspace is fully customizable by the user. Users can rename clusters, drag resources between clusters, reorder clusters, and split or merge clusters. Any such changes are preserved as new resources are saved to the workspace. In this way, the system offers organizational support while allowing users to reshape the workspace to reflect their own understanding of the material~\citep{russell1993cost,zhang2020cognitive}.

Each saved resource retains provenance metadata, including its source-platform icon, the query from which it was saved, and its save timestamp. These cues support later reacquaintance by helping users remember where an item came from and why it may have been relevant~\citep{gomes2022study}. A green \emph{New} badge remains visible until the item is opened or moved, providing a lightweight indication of which saved resources have not yet been directly engaged with after they were added to the workspace. The full workspace state persists across sessions, including saved resources, cluster assignments, cluster labels, and all user modifications, so participants return to the same workspace state in the resumed session.

\subsection{Note-Taking (\textit{NT})}

\begin{figure}
  \centering
  \includegraphics[width=\columnwidth]{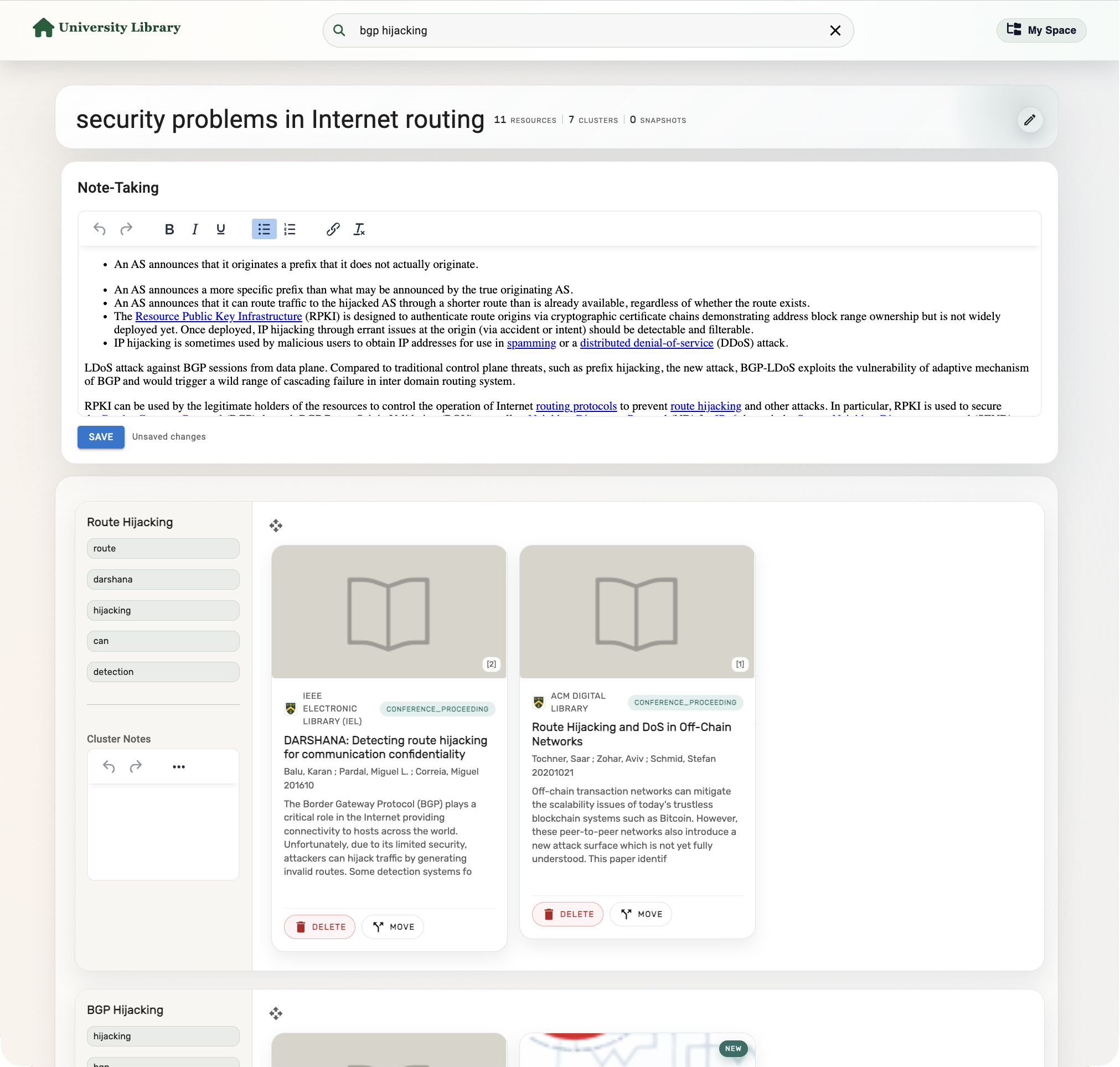}
  \caption{The workspace in the \textit{Note-Taking} (\textit{NT}) condition. Users write their own notes beneath each cluster label; the example shows participant-authored notes covering route hijacking, RPKI, and IP hijacking. A global note area provides space for overall search task notes.}
  \label{fig:wp-nt}
\end{figure}

\textit{Note-Taking} (\textit{NT}) is the human-authored scaffold. Each cluster contains an embedded rich-text editor directly beneath the cluster label (Figure~\ref{fig:wp-nt}), where users can write summaries in their own words, record why resources were saved, note open questions, or leave reminders for later work~\citep{marshall1997annotation,roy2021notehighlight}. Basic formatting is supported, including bulleted lists, bold text, and hyperlinks. In addition to the per-cluster editors, the workspace includes a global note area
for observations that span multiple clusters.

Under the ICAP framework, writing notes that synthesize, reinterpret, or reorganize information reflects \emph{constructive} engagement, because learners must produce their own external representation of what they understand~\citep{chi2014icap}. Prior work has shown that note-taking during search can support downstream writing and learning by helping users encode, organize, and revisit relevant information~\citep{roy2021notehighlight}. All consolidation work in this condition is performed by the learner; the only shared GenAI-authored element visible in both conditions is the cluster label, which is part of the common organizational layer described above. All notes persist across sessions and reappear with the same clusters and saved resources when the learner returns to the search task.

\subsection{Steerable AI Summaries (\textit{STAIS})}
\label{sec:stais}

\begin{figure}
  \centering
  \includegraphics[width=\columnwidth]{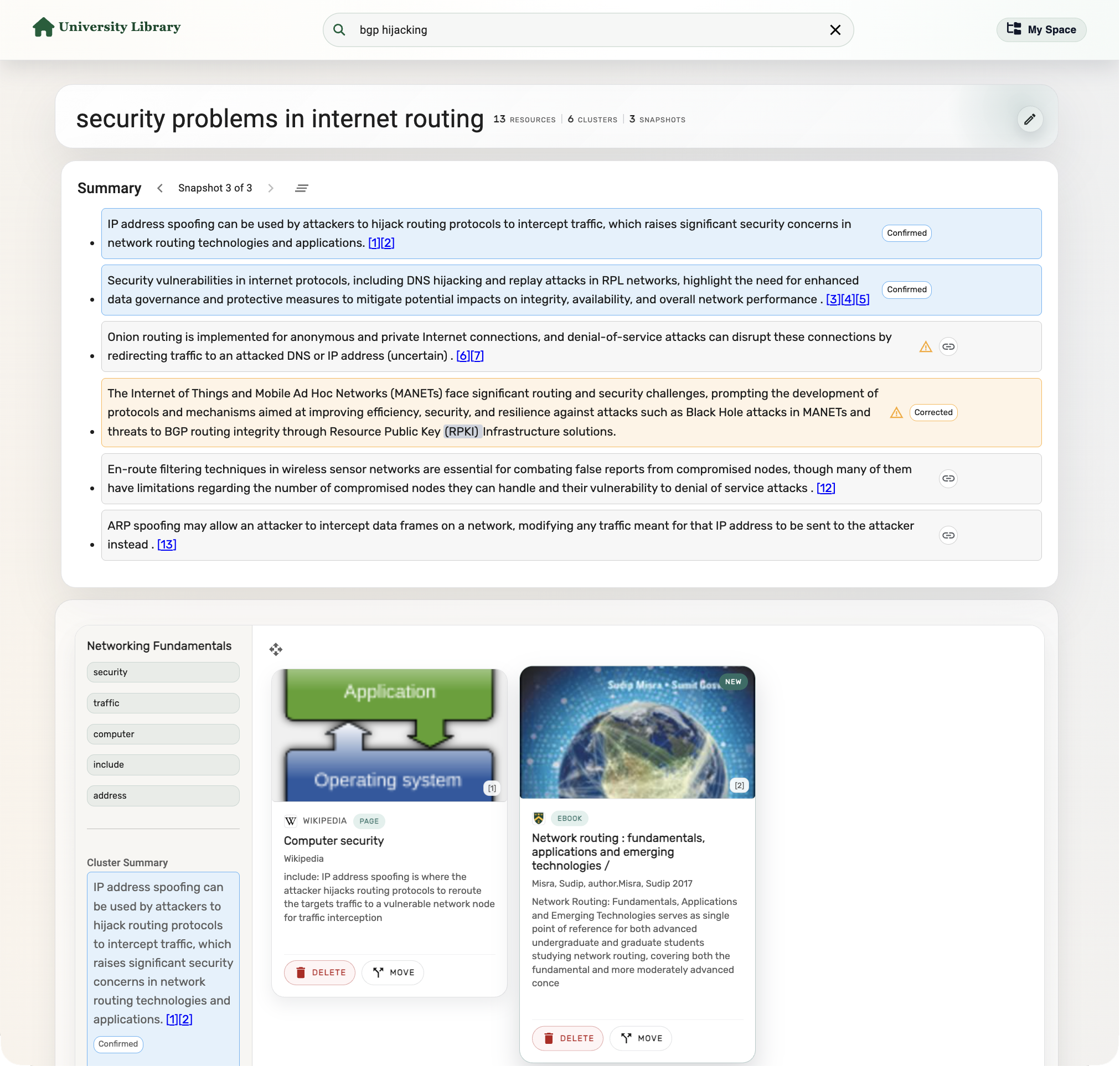}
  \caption{The workspace in the \textit{STAIS} condition. GenAI-generated bullet-point summaries appear with inline citation markers. Each bullet can be edited, and a fact-check mode displays verdict indicators.}
  \label{fig:wp-stais}
\end{figure}

\textit{Steerable AI Summaries} (\textit{STAIS}) is the GenAI-steered scaffold. It provides GenAI summaries that are intended to support active, inspectable, and revisable engagement with the saved materials (Figure~\ref{fig:wp-stais}). Each cluster contains a short bullet-point summary directly beneath its label, and the workspace also includes a summary spanning all clusters. These summaries are generated only from the resources saved within the relevant scope (either the current cluster or the full workspace). Summaries are regenerated automatically whenever the cluster's content changes (e.g., when a resource is added, removed, or moved between clusters), so that the displayed text always reflects the current set of resources.

We chose to design this interface using bullet-point summaries rather than prose paragraphs for two reasons. First, bullets separate individual claims, making the summary easier to scan during reacquaintance. Second, bullets make verification and revision easier because users can inspect or edit one claim at a time~\citep{gao2023citations}. Each bullet ends with inline citation markers (referencing numbered search results within the workspace) linking the claim to its supporting saved resources. When the evidence contains disagreement or tension, the system preserves that distinction in the cited summary rather than collapsing it into a single unsupported statement. As with clustering, summaries and fact-check verdicts are generated by GPT-4o-mini, selected for its balance of cost and energy efficiency, low latency, and adequate quality for evidence-bounded tasks. A larger model might produce more accurate summaries and more reliable fact-check verdicts, but would increase response latency during interactive steering; the observed behaviour should therefore be interpreted as specific to this model family. Prompt design, chunking, and attribution details are reported in Appendix~\ref{app:implementation}.

To reduce passive consumption and preserve user agency, \textit{STAIS} supports three forms of steering~\citep{amershi2019guidelines,shneiderman2020human-centered}. While prior work on steerable GenAI sensemaking has explored visual workspace-based steering~\citep{tang2024steering} and shared human-AI workspaces with iterative refinement~\citep{tang2025ReSPIRE}, \textit{STAIS} differs in that it embeds steering within a cross-session aggregated academic search workspace where summaries are grounded exclusively in user-saved resources and where user corrections persist across sessions and carry forward each time a summary is regenerated. The three types of steering are:

\begin{itemize}
  \item \textbf{Structural steering through reorganization.} Because summaries are generated from the current contents of a cluster, moving resources between clusters, splitting or merging clusters, or renaming clusters changes the evidence context from which the summary is produced.
  
  \item \textbf{Inline bullet correction.} Users can directly revise any summary bullet. Accepted edits are stored as workspace-scoped corrections and are reapplied when the summary is regenerated later.
 
  \item \textbf{Fact-check-driven revision.} A fact-checking mode compares each bullet against its cited evidence and returns one of four verdicts: \emph{Supported}, \emph{Partial}, \emph{Contradicted}, or \emph{Missing Evidence}~\citep{maynez2020faithfulness,huang2024hallucination}. Users can then accept, revise, or dismiss the suggested correction.
\end{itemize}

These mechanisms were designed to shift interaction with GenAI summaries away from simple reading and toward critique, repair, and evidence-based refinement. In ICAP terms, merely reading a generated summary would be considered passive engagement, whereas editing, restructuring, and verifying summary content moves interaction toward constructive, and at times interactive, engagement~\citep{chi2014icap}.

In the \textit{STAIS} condition, the cross-session persistence extends to accepted bullet edits and fact-check-based corrections, which are stored for the user-workspace pair and reapplied whenever a summary is regenerated~\citep{xu2022goldfish,urgo2025search}. This is especially important because summaries are regenerated each time the cluster's content changes; without persistent corrections, a user's prior edits would be lost whenever a new resource is saved or moved.

Because the fact-checking component serves as decision support rather than ground truth, its output is advisory. Users remain responsible for judging whether a flagged claim should be revised and how it should be revised. We therefore treat fact-checking as a support mechanism for reflective review, not as an automated replacement for source inspection. A stratified validation of fact-check accuracy is reported in Appendix~\ref{app:factcheck}.

\section{Evaluation Methodology}
\label{sec:method}
We conducted a controlled cross-session laboratory study to examine how the two cross-session consolidation scaffolds (\textit{NT} and \textit{STAIS}) shape exploratory search, task resumption, continued searching, and post-resumption learning outcomes. Participants completed the same unfamiliar search task in two in-person sessions separated by one week. The two conditions shared the same search interface assignment, information sources, workspace structure, and persistence across sessions. They differed only in the form of consolidation scaffolds available within the workspace: human-authored note-taking (\textit{NT}) or steerable GenAI summaries (\textit{STAIS}). 
The study addressed six research questions:

\textbf{RQ1: Utility.} How do efficiency, effectiveness, and feature use differ between \textit{NT} and \textit{STAIS}?

\textbf{RQ2: Perceived Value.} How do perceived usefulness, ease of use, satisfaction, and user engagement differ between \textit{NT} and \textit{STAIS}?

\textbf{RQ3: Metacognition.} How do \textit{NT} and \textit{STAIS} shape learners' planning, strategy use, monitoring, progress evaluation, and adaptive self-regulation?

\textbf{RQ4: Exploratory Search.} To what extent do \textit{NT} and \textit{STAIS} affect learners' perceptions of how well the workspace supports exploratory search?

\textbf{RQ5: Cross-Session Support.} How well do \textit{NT} and \textit{STAIS} support learners' perceptions of task preparation and task resumption in the resumed session?

\textbf{RQ6: Post-Resumption Learning.} What effects do \textit{NT} and \textit{STAIS} have on self-reported and actual learning outcomes?

\subsection{Study Design}
The study used a controlled laboratory experiment with a between-subjects design~\citep{kelly2009methods}. Participants were assigned to one of two scaffold variants, \textit{NT} or \textit{STAIS}, using block randomization (block size 4) stratified by self-reported gender.

The two-session workflow and the main between-condition comparisons were specified in advance. Correlational analyses, composite scores, and follow-up sensitivity checks were treated as exploratory. The planned workflow consisted of broad exploratory search and workspace consolidation in the initial session, followed by search task resumption, continued searching, and a writing assignment in the resumed session.

\subsection{Measures}
The study used three layers of evidence: interaction logs, self-reported perceptions of value, and a written assessment. Together, they provide a picture of how each scaffold shaped post-resumption work. All survey instruments used 5-point Likert scales.

\textbf{RQ1: Utility.} Utility was operationalized in terms of efficiency, effectiveness, and feature use.

Efficiency was measured by three timing variables, reported chronologically across the session sequence: (1) initial session active search time, derived from system logs; (2) time to task reacquaintance in the resumed session, defined as the interval from when the participant loads the workspace to their first new query submission; and (3) resumed session active search time to task completion, also derived from system logs.

Effectiveness was measured by the precision of saved resources. Two raters, both blind to the condition, independently judged each saved resource on a 3-point scale (relevant, partially relevant, not relevant) to the assigned task scenario; the middle category was collapsed into the relevant category for analysis, as partially relevant search results are valuable for exploratory search. The precision of each participant's saved set was computed as the proportion of saved resources judged relevant.

Feature use was derived from interaction logs. General workspace actions included resources saved, items moved between clusters, and clusters renamed. Condition-specific actions included note creation and editing in \textit{NT}, and summary regeneration, inline bullet correction, and fact-check corrections in \textit{STAIS}. Because each saved resource retained source provenance, we also examined the source composition of the saved evidence base and the proportion of mixed-source clusters as additional indicators of evidence diversity. These same artifacts (authored notes in \textit{NT} and inline bullet corrections in \textit{STAIS}) were also subjected to a secondary qualitative content analysis.

\textbf{RQ2: Perceived Value.} Perceived usefulness and perceived ease of use were measured using TAM2-based scales~\citep{venkatesh2000theoretical,kelly2009methods}. Satisfaction was measured using a post-task satisfaction instrument from prior studies on exploratory search interfaces~\citep{hoeber2022study}. User engagement was measured using the User Engagement Scale Short Form (UES-SF)~\citep{obrien2018practical}, which includes sub-scales of focused attention, perceived usability (reverse coded), aesthetic appeal, and reward factor. 
These instruments were administered after each session.

\textbf{RQ3: Metacognition.} Metacognition was measured using the SRL Perceptions Questionnaire (SPQ)~\citep{urgo2026spq}, a 23-item instrument designed to capture perceptions of self-regulated learning during information seeking. The SPQ covers five sub-scales aligned with the Winne and Hadwin model of self-regulated learning: planning, strategy use, monitoring, evaluating progress, and adapting. The SPQ was administered after each session.

\textbf{RQ4: Exploratory Search.} Learners' perceptions of support provided for exploratory search were measured using a 14-item self-report instrument covering seven dimensions, adapted from prior work~\citep{orin2025thesis}. 
This instrument captures perceived support for discovery, learning, investigation, uncertainty reduction, query reformulation, result examination, and information extraction, which are derived from White \& Roth's model of exploratory search~\citep{white2009exploratory}.
It was administered after each session.

\textbf{RQ5: Cross-Session Support.} Learners' perceptions of how well the workspace supported task preparation and task resumption were measured using a six-item instrument comprising two three-item sub-scales, task preparation (TP) and task resumption (TR), adapted from prior exploratory search studies~\citep{Wu2020understand}. These items were administered immediately after the reacquaintance phase and before participants resumed focused searching, so that responses reflected the perceived ease of resumption itself rather than later task progress.

\textbf{RQ6: Post-Resumption Learning.} Post-resumption learning was measured in two ways. Four self-report learning constructs adapted from prior search-as-learning research~\citep{urgo2025search,urgo2026sal} and cross-session search~\citep{hoeber-islam-2026} were administered before and after each session to track how participants perceived their own learning over time. These included sub-scales of knowledge, interest, certainty, and confidence. At the end of the resumed session, an open-ended writing assignment was administered, from which two count-based outcomes were extracted: \emph{concept count}, the number of accurate, domain-relevant concepts described in the participant's own words; and \emph{connection count}, the number of explicit inter-concept connections among those concepts. 

\subsection{Participants and Task}
\label{sec:participants}

Participants ($n=36$; 19 men, 17 women; mean age $=21.2$, $SD=1.7$) were recruited from the Computer Science Participant Pool at our university and assigned to the two conditions using the stratified block-randomization procedure described in Section~\ref{sec:method}, yielding two equal-sized groups of $n=18$ per condition. An a priori power analysis in G*Power 3~\citep{faul2007gpower} indicated that this allocation provides 80\% power to detect large between-condition effects ($d \geq 0.95$) using a two-tailed independent-samples $t$-test at $\alpha=0.05$~\citep{cohen1992power}; a corresponding sensitivity analysis for Mann-Whitney comparisons indicated detectable effects of approximately $r_{rb} \geq 0.43$. To ensure that observed differences would be attributable to the experimental manipulation rather than prior topical expertise, recruitment was drawn from a second-year course that would be taken prior to courses related to the topic (networking, cybersecurity).  Recruitment materials specified exclusion criteria based on prior exposure to the topic; verification at scheduling time confirmed that no participant had taken any courses related to the topic.

Using simulated work-task methodology~\citep{borlund2003iir}, the task was framed as a realistic academic search problem: a student preparing a short research brief on \emph{security problems in Internet routing} was asked to learn the broad topic, gather evidence-based arguments, and organize materials for later synthesis. The brief was to be grounded in academic literature and verifiable fact-based resources, which is why we paired an academic library with an encyclopedia for this task. The full task description and per-session instructions are provided in Figure~\ref{fig:search_task}.

The domain was selected for three reasons. First, it contains dense technical vocabulary and domain-specific abbreviations (e.g., BGP, RPKI, BGPSEC, AS paths), creating a meaningful gap between surface familiarity and genuine understanding. Second, it has a rich conceptual structure spanning protocol foundations, attack types, mitigation strategies, and governance issues, making it suitable for both breadth-oriented and relation-oriented outcome measures. Third, the topic represents a realistic learning goal for the target participants, yet is sufficiently well-defined to allow prior familiarity to be screened during recruitment.

\begin{figure}
  \centering
  {\setlength{\fboxrule}{0.6pt}\setlength{\fboxsep}{6pt}%
    \fbox{%
      \begin{minipage}{0.95\columnwidth}\small \textbf{Search
        Scenario.} You are a computer science student in a course on computer networking. Your instructor assigns a short research brief on \emph{security problems in Internet routing}. The brief should explain the core problem, describe real-world consequences, and summarize practical mitigation approaches, supported by credible sources.
        \newline \newline
        Your instructor recommends using diverse sources, including academic and professional publications accessed through the university library, along with a general reference source for background. You will use a search interface that aggregates results from the University Academic Library and Wikipedia. Beyond searching, the interface lets you save resources into a workspace and organize what you have found so you can return to it later.
        \newline \newline
        \textbf{Initial Session Instructions.} Repeat the
        scenario in your own words to confirm understanding. Open the
        search interface, create a new workspace for this activity,
        and spend \textbf{10 minutes} searching broadly to learn the
        topic and identify important concepts, examples, and mitigation 
        approaches. Then spend \textbf{5 minutes} saving and organizing your 
        resources in the workspace so your progress will make sense when you 
        return later.
        \newline \newline
        \textbf{Resumed Session Instructions.} Repeat the
        scenario in your own words to confirm understanding. Open the
        search interface, select last week's workspace for this
        activity, and briefly review your saved and organized materials
        to reacquaint yourself. Continue searching to deepen your understanding 
        of the topic, and use the workspace at any time to reorganize, add new 
        resources, or remove items that are no longer relevant.
        \newline \newline
        \textbf{Writing Instructions.} After finishing your search, write a short set of notes in your own words: describe everything that you learned during the search to form the basis for writing the research brief at a later point in time. Base your writing on the evidence you collected and organized in the workspace. As in typical course policies, submissions must be written by you and should not consist of text copied from automated tools.
      \end{minipage}%
    }%
  }
  \caption{Cross-session simulated work task and instructions (networking domain).}
  \label{fig:search_task}
\end{figure}

\subsection{Procedures}
\label{sec:procedures}
The study comprised two in-person sessions separated by one week, consistent with prior studies of cross-session search~\citep{morris2008searchbar,capra2010tools-at-hand,gomes2022study}. Participants received course credit for the initial session, in accordance with the procedures for recruitment out of the participant pool, and a \$25 gift card for completing the resumed session.

In the initial session, participants provided informed consent and completed a pre-study questionnaire. They then watched a video-recorded introduction to the search interface and workspace, followed by a guided training exercise in which they used the live interface to practise querying, viewing item details, saving resources, organizing clusters, and using the key features that are unique to their assigned interface. 

Participants then undertook the initial search session, which consisted of 10 minutes of exploratory searching and 5 minutes of workspace consolidation. Participants were notified at the 10-minute mark that they had 5 minutes left to organize their saved resources. After the task, they completed a set of post-task questionnaires and were reminded of the schedule for the second session.

The resumed session began with a pre-session questionnaire and a brief video-recorded refresher on the assigned interface, after which participants could try out the interface features to reacquaint themselves before resuming the task. Participants reopened the workspace they had created in the initial session and completed a reacquaintance phase. Immediately after reacquaintance, and before continuing the task, they completed the task preparation and task resumption survey. Participants then continued searching to deepen their understanding of the same topic while further revising their workspace. After the participants decided they had found enough information for the assigned task, they completed the remaining post-task questionnaires. 

The final activity of the study requested that participants write their own notes about what they had learned. During the writing assignment, the screen was split so that the workspace remained visible beside the writing area; participants could therefore consult their saved resources, cluster structure, and the consolidation artifacts available within their assigned search interface.

All sessions were conducted in the same lab, on the same desktop workstation, with a fixed browser window, and were screen- and audio-recorded for later coding. Eye-tracking and retrospective think-aloud data were also collected during the study. Because of the richness and complexity of those data, they are outside the scope of the present publication.

The full study protocol, including the consent form, participant recruitment, compensation methods, and all data collection methods and instruments, was reviewed and approved by our university's Research Ethics Board prior to data collection.

\subsection{Data Analysis}
\label{sec:data-analysis}
Baseline equivalence between the two conditions with respect to demographics and pre-study questionnaire data was verified (Table~\ref{tab:equivalence}); no condition differences emerged, so all subsequent between-condition comparisons proceed without covarying these variables. Normality was assessed using the Shapiro-Wilk test, and homogeneity of variance was checked where appropriate to guide test selection. The significance threshold was set at $\alpha=0.05$.

\begin{table}[!htbp]
  \centering
  \caption{Baseline equivalence checks between \textit{NT} and \textit{STAIS}. Means ($SD$) per condition with the corresponding test statistic. Baseline domain familiarity was measured on a 5-point Likert scale (1 = none, 5 = expert).}
  \label{tab:equivalence}
  \small
  \begin{tabular}{lccl}
    \toprule
    Variable & \textit{NT} & \textit{STAIS} & Test \\
    \midrule
    Age (years)                  & 21.3 (1.8)  & 21.1 (1.6)  & $t(34) = 0.42$, $p = .677$ \\
    Gender (women / men)         & 8 / 10      & 9 / 9       & $\chi^2(1) = 0.11$, $p = .740$ \\
    Baseline domain familiarity  & 1.78 (0.55) & 1.83 (0.61) & $t(34) = 0.38$, $p = .706$ \\
    \bottomrule
  \end{tabular}
\end{table}

Each of the survey instruments used in this study consisted of one or more groups of questions that were combined into composite measures.
Internal consistency was computed across the items of each composite using Cronbach's $\alpha$, which are reported in Table~\ref{tab:reliability}. In all cases, the internal consistency met or exceeded the ``acceptable'' threshold ($\alpha \geq .70$;~\citealt{nunnally1994psychometric}). The composite was created as the mean value of the specific survey items.

\begin{table}[!htbp]
  \centering
  \caption{Internal consistency (Cronbach's $\alpha$) of composite measures across administrations. Conventional interpretations: acceptable ($\alpha \geq .70$), good ($\alpha \geq .80$), excellent ($\alpha \geq .90$).}
  \label{tab:reliability}
  \small
  \begin{tabular}{ll}
    \toprule
    Composite & $\alpha$ \\
    \midrule
    Perceived usefulness, ease of use, satisfaction (3 sub-scales)      & 0.79 - 0.84       \\
    User engagement (UES-SF: 4 sub-scales)                   & 0.72 - 0.86  \\
    Perception of self-regulated learning (SPQ: 5 sub-scales)                      & 0.73 - 0.88  \\
    Cross-session support (2 sub-scales)    & 0.76 - 0.80  \\
    Self-reported learning (4 sub-scales)   & 0.71 - 0.82  \\
    \bottomrule
  \end{tabular}
\end{table}

Composite survey scores were treated as approximately continuous. When normality assumptions were satisfied, between-condition comparisons used independent-samples $t$-tests, with Cohen's $d$ reported as the effect-size measure when the comparison reached statistical significance. Count-based variables, including feature-use counts and writing assignment concept and connection counts, were analyzed using Mann-Whitney tests, with rank-biserial correlation ($r_{rb}$) reported as the corresponding effect-size measure when the comparison reached statistical significance. Following common practice in human-centred information-retrieval studies that contrast two conditions per research question, no family-wise correction was applied; instead, exact $p$-values are reported throughout, including for non-significant comparisons.

For the four self-report learning constructs, cross-session change was computed as the post-minus-pre difference for each session, following the analytical approach used with a similar instrument in prior research~\citep{hoeber-islam-2026}. Statistical tests were conducted on the cross-session change with respect to the two conditions. 

The results of the writing assignment were scored independently by two raters using a count-based coding scheme grounded in the Wilson and Wilson~\citep{wilson2013assessment} dimensions and operationalized following prior search-as-learning work~\citep{urgo2026sal}. Each rater extracted concept and connection counts from each participant's writing. 

To characterize the nature of engagement within each condition, two independent coders categorized all authored notes (\textit{NT}) and all inline bullet corrections (\textit{STAIS}) using condition-specific five-category taxonomies ordered by increasing cognitive depth. For \textit{STAIS} corrections, the categories were factual error (model confabulation or user-introduced), source misalignment, precision refinement, interpretive reframing, and integrative elaboration. For \textit{NT}, the categories were verbatim extraction, paraphrased summary, cross-source synthesis, evaluative annotation, and organizational planning. Because the two taxonomies were not one-to-one equivalents, this analysis was used to characterize engagement within each condition, not to claim direct equivalence across conditions. Within-condition change between the initial and resumed sessions was assessed using Wilcoxon signed-rank tests on raw counts.

All hand-coded measures were independently rated by two raters before analysis, with disagreements resolved through discussion and consensus. Agreement on the relevance judgments used to compute saved-resource precision was ``substantial''~\citep[$\kappa=0.81$;][]{landis1977kappa}; agreement on the concept and connection counts from the writing assignments were ``substantial'' and ``excellent'', respectively~\citep[concept count ICC $=0.82$, connection count ICC $=0.85$;][]{landis1977kappa,koo1996icc}; and agreement on the qualitative coding of authored notes (\textit{NT}) and inline corrections (\textit{STAIS}) for the secondary content analysis was ``substantial''~\citep[$\kappa=0.79$;][]{landis1977kappa}.

\section{Results}
\label{sec:results}

We report between-condition comparisons for RQ1-RQ6. 
In each results table, exact $p$-values are reported for non-significant outcomes, and boundary $p$-values along with effect sizes are reported for significant outcomes. When data is not available for a given measure for a given condition in the study (due to the nature of the interface or the design of the study), a dash (--) is used to signify this, and the space for the results of the statistical analysis is left blank. Bold table entries mark statistically significant results ($p < 0.05$); a dagger ($\dagger$) following an effect-size value marks a medium effect ($0.50 \leq |d| < 0.80$ or $0.30 \leq r_{rb} < 0.50$), and an asterisk ($*$) marks a large effect ($|d| \geq 0.80$ or $r_{rb} \geq 0.50$).

\subsection{RQ1: Utility}

All utility and interaction data for RQ1 are reported in Table~\ref{tab:utility}. In the initial session, task efficiency, precision of saved resources, and general interaction patterns (items saved and items moved) did not differ significantly between the two conditions; the near-identical times on task reflect the fixed allocation of 15 minutes. Relative to the shared baseline interface on which both conditions were built, \textit{NT} and \textit{STAIS} diverged only on the condition-specific actions each added: \textit{Note-Taking} participants created and edited notes, whereas \textit{STAIS} participants regenerated summaries, made inline bullet corrections, and used fact-checking. Both conditions showed active workspace engagement during the initial session.

In the resumed session, \textit{NT} and \textit{STAIS} differed on three utility measures: \textit{STAIS} showed a shorter time to reacquaintance, higher precision of saved resources, and more items moved between clusters. While the participants who used \textit{NT} created and edited fewer notes in the resumed session, those who used \textit{STAIS} increased their use of the four steering actions. In one case, this increase was statistically significant: inline bullet corrections (Wilcoxon signed-rank $Z = 3.52$, $p < 0.001$). 

Further analysis of the nature of the interaction with the interfaces is provided in the sub-sections that follow. These focus on consolidation artifacts created in the two interfaces, and the types of sources saved and used.

\begin{table*}[!htbp]
  \caption{Mean values (standard deviation) for utility measures collected during the \textit{initial} and \textit{resumed} sessions.
    }
  \label{tab:utility}
  \resizebox{\linewidth}{!}{%
  \small
  \begin{tabular}{l cc l cc l}
    \toprule
    & \multicolumn{3}{c}{\textit{Initial}} & \multicolumn{3}{c}{\textit{Resumed}} \\
    \cmidrule(lr){2-4} \cmidrule(lr){5-7}
    Measure & \textit{NT} & \textit{STAIS} & Statistic & \textit{NT} & \textit{STAIS} & Statistic \\
    \midrule
    Time on task (min)        & 14:55 (0:20) & 14:50 (0:25) & $t(34)=0.66$, $p=0.51$ & 18:10 (2:15) & 16:40 (2:40) & $t(34)=1.82$, $p=0.08$ \\
    Time to reacquaintance    & --           & --           &                      & 1:35 (0:45)  & \textbf{0:55 (0:30)}  & $\mathbf{t(34)=3.13}$, $\mathbf{p<0.01}$, $d=1.04*$ \\
    Precision                 & 0.74 (0.11)  & 0.76 (0.10)  & $t(34)=0.57$, $p=0.57$ & 0.74 (0.10)  & \textbf{0.83 (0.09)}  & $\mathbf{t(34)=2.84}$, $\mathbf{p<0.01}$, $d=0.95*$ \\
    \midrule
    Items saved               & 5.1 (1.5)    & 5.3 (1.6)    & $U=155$, $p=0.84$ & 4.8 (1.4) & 5.7 (1.5) & $U=104$, $p=0.07$ \\
    Items moved               & 3.8 (1.5)    & 4.2 (1.6)    & $U=147$, $p=0.65$ & 3.5 (1.8) & \textbf{5.8 (2.1)} & $\mathbf{U=78}$, $\mathbf{p<0.01}$, $r_{rb}=0.52*$ \\
    \midrule
    Notes created / edited    & 4.2 (1.7)    & --           &    & 2.8 (1.3)    & --           &    \\
    Summaries regenerated     & --           & 3.1 (1.4)    &    & --           & 4.7 (1.8)    &    \\
    Inline bullet corrections & --           & 2.9 (1.6)    &    & --           & 5.4 (2.3)    &    \\
    Fact-check activations    & --           & 4.2 (2.0)    &    & --           & 7.8 (2.6)    &    \\
    Fact-check corrections    & --           & 1.2 (0.8)    &    & --           & 2.8 (1.4)    &    \\
    \bottomrule
  \end{tabular}}
\end{table*}

\subsubsection{Consolidation Artifacts}

The results of a content analysis of the consolidation artifacts created with the two interfaces are reported in Table~\ref{tab:content-analysis}. For each condition, the five coding categories are ordered from lower-order (reproductive) activities to higher-order (generative or interpretive) activities.

For participants who used \textit{NT}, there was a statistically significant reduction in the number of lower-order activities, with no substantive change in the number of higher-order activities undertaken. For participants who used \textit{STAIS}, there was a statistically significant increase in one of the low-order activities (factual error) and a statistically significant increase in all high-order activities.

Both conditions showed evidence of engaging in more cognitively demanding activities from the initial to the resumed session, but reached these through different routes. The \textit{NT} participants shed lower-order activities, resulting in an increase in the proportion of higher-order activities being undertaken, whereas the \textit{STAIS} participants added more higher-order corrections on top of a continued need to correct factual errors produced by the GenAI.

\begin{table*}[!htbp]
  \caption{Mean counts (standard deviation) for the content analysis of the authored notes (\textit{NT}) and inline corrections (\textit{STAIS}) collected during the \textit{initial} and \textit{resumed} sessions. 
  }
  \label{tab:content-analysis}
  \resizebox{\linewidth}{!}{%
  \small
  \begin{tabular}{llp{7cm}ccl}
    \toprule
    & Category & Description & Initial & Resumed & Statistic \\
    \midrule
    \multirow{6}{*}{\rotatebox[origin=c]{90}{\textit{NT}}}
      & \multicolumn{5}{l}{\textit{Lower-order activities}} \\
      & \quad Verbatim extraction     & Direct quotation or near-copy of source text     & \textbf{1.2 (0.8)} & 0.3 (0.3) & $\mathbf{Z=3.06}$, $\mathbf{p<0.01}$, $r_{rb}=0.51*$ \\
      & \quad Paraphrased summary     & Source content restated in own words             & \textbf{1.4 (0.9)} & 0.6 (0.5) & $\mathbf{Z=2.89}$, $\mathbf{p<0.01}$, $r_{rb}=0.48\dagger$ \\
      \cmidrule{2-6}
      & \multicolumn{5}{l}{\textit{Higher-order activities}} \\
      & \quad Cross-source synthesis  & Note connecting ideas from multiple sources      & 0.7 (0.5) & 0.9 (0.6) & $Z=1.34$, $p=0.18$ \\
      & \quad Evaluative annotation   & Personal assessment or critique of source content & 0.5 (0.4) & 0.7 (0.5) & $Z=1.51$, $p=0.13$ \\
      & \quad Organizational planning & Note about task structure, gaps, or next steps    & 0.4 (0.3) & 0.3 (0.3) & $Z=0.82$, $p=0.41$ \\
    \midrule
    \multirow{6}{*}{\rotatebox[origin=c]{90}{\textit{STAIS}}}
      & \multicolumn{5}{l}{\textit{Lower-order activities}} \\
      & \quad Factual error          & Corrected claim with no grounding in cited source                   & 1.2 (0.8) & \textbf{2.1 (1.1)} & $\mathbf{Z=2.83}$, $\mathbf{p<0.01}$, $r_{rb}=0.47\dagger$ \\
      & \quad Source misalignment    & Corrected divergence from cited source                          & 0.5 (0.4) & 0.8 (0.6) & $Z=1.72$, $p=0.09$ \\
      \cmidrule{2-6}
      & \multicolumn{5}{l}{\textit{Higher-order activities}} \\
      & \quad Precision refinement   & Overly broad claim narrowed for specificity                & 0.5 (0.4) & \textbf{0.9 (0.6)} & $\mathbf{Z=2.15}$, $\mathbf{p<0.05}$, $r_{rb}=0.36\dagger$ \\
      & \quad Interpretive reframing & Rewritten to foreground participant's understanding & 0.4 (0.4) & \textbf{1.0 (0.7)} & $\mathbf{Z=2.67}$, $\mathbf{p<0.01}$, $r_{rb}=0.45\dagger$ \\
      & \quad Integrative elaboration & Rewritten to incorporate connections across sources    & 0.3 (0.3) & \textbf{0.6 (0.5)} & $\mathbf{Z=1.98}$, $\mathbf{p<0.05}$, $r_{rb}=0.33\dagger$ \\
    \bottomrule
  \end{tabular}}
\end{table*}

\subsubsection{Source Type Composition}
\label{sec:source-composition}

The results of analyzing the source types among the saved resources and corresponding clusters in the workspace are reported in Table~\ref{tab:source-type}. Because both conditions drew from the same pool of source information (an academic library and Wikipedia), any difference in the composition of the saved evidence reflects what the participants chose to save rather than what was available. In the initial session, participants who used \textit{NT} and \textit{STAIS} saved resources from the two sources in similar proportions.
In the resumed session, there was a statistically significant increase in the number of resources saved from Wikipedia for the participants who used \textit{STAIS}, which resulted in a statistically significant increase in the proportion of search results saved from the library for the participants who used \textit{NT}. This increase in saved Wikipedia resources also resulted in more mixed-source clusters for the participants who used \textit{STAIS}, which shows that the Wikipedia resources were integrated into the clusters rather than isolated into their own clusters.

One interpretation is that summarization and fact-checking in \textit{STAIS} surfaced specific claims whose verification was easier to ground in a broad reference source, encouraging participants to add encyclopedic evidence alongside academic material. \textit{NT} participants, whose consolidation activity centred on authoring their own notes, appear to have leaned more heavily on the Library as the primary source for those notes.

\begin{table}[!htbp]
  \centering
  \caption{Mean values (standard deviation) for source type composition of saved resources collected during the \textit{initial} and \textit{resumed} sessions.
    }
  \label{tab:source-type}
  \small
  \resizebox{\columnwidth}{!}{%
  \begin{tabular}{l cc l cc l}
    \toprule
    & \multicolumn{3}{c}{\textit{Initial}} & \multicolumn{3}{c}{\textit{Resumed}} \\
    \cmidrule(lr){2-4} \cmidrule(lr){5-7}
    Measure & \textit{NT} & \textit{STAIS} & Statistic & \textit{NT} & \textit{STAIS} & Statistic \\
    \midrule
    Library saves         & 2.9 (1.1) & 3.0 (1.2) & $U=154$, $p=0.82$ & 3.0 (1.1) & 2.8 (1.1) & $U=142$, $p=0.56$ \\
    Wikipedia saves       & 2.2 (1.0) & 2.3 (0.9) & $U=156$, $p=0.86$ & 1.8 (0.8) & \textbf{2.9 (1.0)} & $\mathbf{U=82}$, $\mathbf{p<0.01}$, $r_{rb}=0.49\dagger$ \\
    Total saves           & 5.1 (1.5) & 5.3 (1.6) & $U=155$, $p=0.84$ & 4.8 (1.4) & 5.7 (1.5) & $U=104$, $p=0.07$ \\
    \midrule
    Library proportion    & 0.57 (0.14) & 0.57 (0.13) & $t(34)=0.10$, $p=0.92$     & \textbf{0.63 (0.14)} & 0.49 (0.13) & $\mathbf{t(34)=3.11}$, $\mathbf{p<0.01}$, $d=1.04*$ \\
    Mixed-source clusters & 0.50 (0.22) & 0.53 (0.20) & $t(34)=0.43$, $p=0.67$      & 0.42 (0.20)          & \textbf{0.62 (0.19)} & $\mathbf{t(34)=3.08}$, $\mathbf{p<0.01}$, $d=1.03*$ \\
    \bottomrule
  \end{tabular}}
\end{table}

\subsection{RQ2: Perceived Value}

Perceived value data are reported in Table~\ref{tab:perceived}. There were no statistically significant differences between the conditions after the initial session. One interpretation is that both conditions were built on the same underlying search workspace, and participants had not yet experienced how the consolidation affordances that differentiate the two interface conditions support cross-session searching.

In the resumed session, perceived usefulness, perceived satisfaction, and the reward factor sub-scale of user engagement were higher for \textit{STAIS}, with large effect sizes. This result is consistent with the utility advantages that are at the foundation of \textit{STAIS}: having system-generated summaries already available upon return appears to have made the resumed work feel more useful, satisfying, and rewarding. The dimensions tied to the shared baseline interface, including ease of use and the remaining user engagement sub-scales (focused attention, perceived usability, and aesthetic appeal), did not differ between conditions, consistent with the fact that both conditions used identical search, clustering, and organizational features.

\begin{table*}[!htbp]
  \caption{Mean values (standard deviation) for perceived usefulness, ease of use, satisfaction, and the four aspects of user engagement (1--5 Likert scales) collected during the \textit{initial} and \textit{resumed} sessions. 
  }
  \label{tab:perceived}
  \resizebox{\linewidth}{!}{%
  \small
  \begin{tabular}{l cc l cc l}
    \toprule
    & \multicolumn{3}{c}{\textit{Initial}} & \multicolumn{3}{c}{\textit{Resumed}} \\
    \cmidrule(lr){2-4} \cmidrule(lr){5-7}
    Scale & \textit{NT} & \textit{STAIS} & Independent $t$-test & \textit{NT} & \textit{STAIS} & Independent $t$-test \\
    \midrule
    Perceived usefulness  & 3.61 (0.68) & 3.94 (0.62) & $t(34)=1.52$, $p=0.137$ & 3.72 (0.60) & \textbf{4.36 (0.57)} & $\mathbf{t(34)=3.28}$, $\mathbf{p<0.01}$, $d=1.09*$ \\
    Perceived ease of use & 4.17 (0.55) & 3.89 (0.63) & $t(34)=1.42$, $p=0.165$ & 4.22 (0.52) & 4.06 (0.58) & $t(34)=0.87$, $p=0.390$ \\
    Perceived satisfaction & 3.67 (0.71) & 3.89 (0.66) & $t(34)=0.96$, $p=0.342$ & 3.72 (0.58) & \textbf{4.33 (0.52)} & $\mathbf{t(34)=3.32}$, $\mathbf{p<0.01}$, $d=1.11*$ \\
    \midrule
    Focused attention     & 3.56 (0.75) & 3.50 (0.72) & $t(34)=0.24$, $p=0.808$ & 3.78 (0.70) & 3.61 (0.68) & $t(34)=0.74$, $p=0.465$ \\
    Perceived usability   & 3.56 (0.72) & 3.72 (0.68) & $t(34)=0.69$, $p=0.498$ & 3.67 (0.69) & 3.89 (0.63) & $t(34)=1.00$, $p=0.325$ \\
    Aesthetic appeal      & 3.78 (0.60) & 3.83 (0.58) & $t(34)=0.25$, $p=0.801$ & 3.83 (0.57) & 3.89 (0.55) & $t(34)=0.32$, $p=0.750$ \\
    Reward factor         & 3.44 (0.80) & 3.67 (0.75) & $t(34)=0.89$, $p=0.380$ & 3.44 (0.72) & \textbf{4.17 (0.62)} & $\mathbf{t(34)=3.26}$, $\mathbf{p<0.01}$, $d=1.09*$ \\
    \bottomrule
  \end{tabular}}
\end{table*}

\subsection{RQ3: Metacognition}

Results from administering the SRL Perceptions Questionnaire (SPQ) are reported in Table~\ref{tab:metacog}. During the initial search session, participants who used \textit{STAIS} reported higher levels of strategy use. This continued in the resumed session, where they also reported higher levels of planning and evaluating progress. Those who used \textit{NT} reported higher levels of adaptation in the resumed session only.

A plausible interpretation is that GenAI summaries externalized strategic options by making structure visible through cluster summaries, source-linked bullets, and fact-checking cues, thereby lowering the cost of choosing among strategies and judging whether progress on the task was on track. By contrast, the activity of writing one's own notes in the \textit{NT} condition required participants to construct and maintain their own external organization of the domain, giving them a sense of being able to adapt to new information as it was discovered.

\begin{table*}[!htbp]
  \caption{Mean values (standard deviation) for metacognitive subscale scores (1--5 Likert scales) collected during the \textit{initial} and \textit{resumed} sessions. 
  }
  \label{tab:metacog}
  \resizebox{\linewidth}{!}{%
  \small
  \begin{tabular}{l cc l cc l}
    \toprule
    & \multicolumn{3}{c}{\textit{Initial}} & \multicolumn{3}{c}{\textit{Resumed}} \\
    \cmidrule(lr){2-4} \cmidrule(lr){5-7}
    Subscale & \textit{NT} & \textit{STAIS} & Independent $t$-test & \textit{NT} & \textit{STAIS} & Independent $t$-test \\
    \midrule
    Planning            & 3.50 (0.67) & 3.72 (0.60)          & $t(34)=1.04$, $p=0.307$ & 3.56 (0.60) & \textbf{4.08 (0.52)} & $\mathbf{t(34)=2.78}$, $\mathbf{p<0.01}$, $d=0.93*$ \\
    Strategy Use        & 3.35 (0.67) & \textbf{3.81 (0.63)} & $\mathbf{t(34)=2.12}$, $\mathbf{p<0.05}$, $d=0.71\dagger$ & 3.44 (0.68) & \textbf{4.11 (0.57)} & $\mathbf{t(34)=3.20}$, $\mathbf{p<0.01}$, $d=1.07*$ \\
    Monitoring          & 3.44 (0.75) & 3.56 (0.70)          & $t(34)=0.50$, $p=0.623$ & 3.56 (0.72) & 3.67 (0.68)          & $t(34)=0.47$, $p=0.640$ \\
    Evaluating Progress & 3.39 (0.73) & 3.50 (0.68)          & $t(34)=0.47$, $p=0.643$ & 3.44 (0.58) & \textbf{4.00 (0.57)} & $\mathbf{t(34)=2.92}$, $\mathbf{p<0.01}$, $d=0.97*$ \\
    Adapting            & 3.61 (0.69) & 3.56 (0.71)          & $t(34)=0.21$, $p=0.832$ & \textbf{4.08 (0.58)} & 3.53 (0.56) & $\mathbf{t(34)=2.89}$, $\mathbf{p<0.01}$, $d=0.96*$ \\
    \bottomrule
  \end{tabular}}
\end{table*}

\subsection{RQ4: Exploratory Search}

Perceptions of how well the search interfaces supported different aspects of exploratory search are reported in Table~\ref{tab:exploratory}. After the initial search session, all seven dimensions were statistically indistinguishable between conditions. After the resumed session, two dimensions differed between the conditions, in opposite directions: perceived support for \emph{uncertainty reduction} was higher for \textit{NT}, and perceived support for \emph{information extraction} was higher for \textit{STAIS}. These differences track the consolidation mechanisms themselves. The active engagement of authoring notes requires \textit{NT} participants to commit to explicit formulations of what they understand, an activity that is well-suited to reducing a searcher's perceived uncertainty about the topic. \textit{STAIS}, by contrast, automates much of the extractive work through GenAI summaries, source-linked bullets, and fact-checking, which made information extraction feel easier.

\begin{table*}[!htbp]
  \caption{Mean values (standard deviation) for perceived support for exploratory search (1--5 Likert scales) collected during the \textit{initial} and \textit{resumed} sessions. 
  }
  \label{tab:exploratory}
  \resizebox{\linewidth}{!}{%
  \small
  \begin{tabular}{l cc l cc l}
    \toprule
    & \multicolumn{3}{c}{\textit{Initial}} & \multicolumn{3}{c}{\textit{Resumed}} \\
    \cmidrule(lr){2-4} \cmidrule(lr){5-7}
    Dimension & \textit{NT} & \textit{STAIS} & Independent $t$-test & \textit{NT} & \textit{STAIS} & Independent $t$-test \\
    \midrule
    Discovery              & 3.56 (0.68) & 3.61 (0.65) & $t(34)=0.23$, $p=0.823$ & 3.72 (0.65) & 3.83 (0.62) & $t(34)=0.52$, $p=0.607$ \\
    Learning               & 3.56 (0.68) & 3.44 (0.70) & $t(34)=0.52$, $p=0.605$ & 3.83 (0.62) & 3.67 (0.65) & $t(34)=0.76$, $p=0.455$ \\
    Investigation          & 3.61 (0.65) & 3.67 (0.62) & $t(34)=0.28$, $p=0.779$ & 3.83 (0.62) & 3.94 (0.58) & $t(34)=0.55$, $p=0.586$ \\
    Uncertainty reduction  & 3.61 (0.70) & 3.44 (0.72) & $t(34)=0.72$, $p=0.478$ & \textbf{4.14 (0.60)} & 3.50 (0.62) & $\mathbf{t(34)=3.15}$, $\mathbf{p<0.01}$, $d=1.05*$ \\
    Query reformulation    & 3.67 (0.64) & 3.72 (0.61) & $t(34)=0.24$, $p=0.812$ & 3.78 (0.62) & 3.83 (0.59) & $t(34)=0.25$, $p=0.806$ \\
    Result examination     & 3.67 (0.66) & 3.83 (0.63) & $t(34)=0.74$, $p=0.462$ & 3.83 (0.63) & 4.00 (0.60) & $t(34)=0.83$, $p=0.413$ \\
    Information extraction & 3.44 (0.72) & 3.67 (0.68) & $t(34)=0.99$, $p=0.331$ & 3.50 (0.62) & \textbf{4.08 (0.55)} & $\mathbf{t(34)=2.97}$, $\mathbf{p<0.01}$, $d=0.99*$ \\
    \bottomrule
  \end{tabular}}
\end{table*}

\subsection{RQ5: Cross-Session Support}

Table~\ref{tab:resumption} reports the participants' perceived support for task preparation and task resumption, which was collected immediately after the reacquaintance period. Perceptions of support for task preparation did not differ significantly between the two interface conditions, indicating that participants in both conditions felt comparably ready to resume work once they had reacquainted themselves with the workspace. Perceptions of support for task resumption were significantly higher for \textit{STAIS}. 

A plausible explanation lies in the scaffolds themselves. Returning \textit{STAIS} participants were presented with well-written GenAI summaries that externalized the structure of their prior work, providing a ready-made entry point into the resumed task. By contrast, returning \textit{NT} participants were presented with their previously-prepared notes, which may have been more difficult to consume due to the lack of an imposed structure and the potential for incomplete writing. The GenAI-generated summaries thus lowered the perceived effort required to pick up where the work had been left off.

\begin{table}[!htbp]
  \centering
  \caption{Mean values (standard deviation) for perceived cross-session search support (1--5 Likert scales) collected at the start of the \textit{resumed} session, after the reacquaintance period. 
  }
  \label{tab:resumption}
  \small
  \begin{tabular}{lccl}
    \toprule
    Scale & \textit{NT} & \textit{STAIS} & Independent $t$-test \\
    \midrule
    Task preparation & 3.72 (0.63) & 3.89 (0.58)         & $t(34)=0.84$, $p=0.406$ \\
    Task resumption  & 3.56 (0.70) & \textbf{4.17 (0.55)} & $\mathbf{t(34)=2.91}$, $\mathbf{p<0.01}$, $d=0.97*$ \\
    \bottomrule
  \end{tabular}
\end{table}

\subsection{RQ6: Post-Resumption Learning}

We report the self-report learning measures first, as they were collected before and after each session, followed by the writing assignment outcomes collected at the end of the resumed session.

\subsubsection{Self-Report Learning}

The self-report learning measured across the two sessions is reported in Table~\ref{tab:learning}. The general pattern across both conditions is an increase after the initial search session, a decrease during the gap, but still above the first measurement prior to the initial search, and another increase during the resumed search session to the highest levels measured. This trajectory is consistent with prior work on search interfaces to support cross-session searching~\citep{hoeber-islam-2026}. The one outlier to this pattern was the confidence measure during the initial search, for the participants who were in the \textit{STAIS} condition.

Between the conditions, participants who used \textit{STAIS} reported greater increases in knowledge and interest, in both the initial and resumed search sessions. Participants who used \textit{NT} reported greater increases in certainty, in both the initial and resumed sessions; they also reported greater increases in confidence in the initial session only. The GenAI summaries in \textit{STAIS} appeared to provide searchers with a sense of increasing their knowledge and interest, perhaps due to the automatic extraction of salient aspects of the saved resources. However, the ability to make corrections, as well as the known issues with confabulations, may have left the participants looking for errors, reducing their certainty and confidence. For those who used \textit{NT}, the process of writing one's own notes may have provided the participants with a sense of certainty and confidence in what they had found. Of note, in the resumed session, there were no differences in confidence among the participants; perhaps those who used \textit{STAIS} were able to gain more confidence as they found more information and added their own learning into the generated bullet points, arriving at a similar increase as those who wrote their own notes using \textit{NT}.

\begin{table*}[!htbp]
  \centering
  \caption{Mean values (standard deviation) for self-report learning measures collected before and after the \textit{initial} and \textit{resumed} sessions. 
  }
  \label{tab:learning}
  \resizebox{\linewidth}{!}{%
  \begin{tabular}{l|l|ccc|l|ccc|l}
    \toprule
    & & \multicolumn{4}{c|}{\textbf{Initial}} & \multicolumn{4}{c}{\textbf{Resumed}} \\
    \cline{3-10}
    \textbf{Measure} & \textbf{Condition} & Prior & Post & $\Delta$ & \textbf{Independent $t$-test} & Prior & Post & $\Delta$ & \textbf{Independent $t$-test} \\
    \midrule
    \multirow{2}{*}{Knowledge}
      & \textit{NT}    & 2.45 (0.54) & 2.86 (0.61)          & $+0.41$ & \multirow{2}{*}{$\mathbf{t(34)=2.82}$, $\mathbf{p<0.01}$, $d=0.94*$}  & 2.55 (0.57) & 3.08 (0.62)          & $+0.53$ & \multirow{2}{*}{$\mathbf{t(34)=3.64}$, $\mathbf{p<0.001}$, $d=1.21*$} \\
      & \textit{STAIS} & 2.38 (0.52) & \textbf{3.42 (0.58)} & $+1.04$ &                                                                               & 3.05 (0.62) & \textbf{3.82 (0.60)} & $+0.77$ & \\
    \midrule
    \multirow{2}{*}{Interest}
      & \textit{NT}    & 2.62 (0.58) & 3.10 (0.64)          & $+0.48$ & \multirow{2}{*}{$\mathbf{t(34)=2.29}$, $\mathbf{p<0.05}$, $d=0.76\dagger$}  & 2.88 (0.62) & 3.25 (0.65)          & $+0.37$ & \multirow{2}{*}{$\mathbf{t(34)=2.54}$, $\mathbf{p<0.05}$, $d=0.85*$} \\
      & \textit{STAIS} & 2.55 (0.61) & \textbf{3.58 (0.62)} & $+1.03$ &                                                                               & 2.95 (0.64) & \textbf{3.78 (0.60)} & $+0.83$ & \\
    \midrule
    \multirow{2}{*}{Certainty}
      & \textit{NT}    & 2.42 (0.51) & \textbf{3.28 (0.58)} & $+0.86$ & \multirow{2}{*}{$\mathbf{t(34)=2.85}$, $\mathbf{p<0.01}$, $d=0.95*$}  & 2.92 (0.55) & \textbf{3.52 (0.60)} & $+0.60$ & \multirow{2}{*}{$\mathbf{t(34)=2.07}$, $\mathbf{p<0.05}$, $d=0.69\dagger$} \\
      & \textit{STAIS} & 2.45 (0.53) & 2.72 (0.60)          & $+0.27$ &                                                                               & 2.64 (0.58) & 3.10 (0.62)          & $+0.46$ & \\
    \midrule
    \multirow{2}{*}{Confidence}
      & \textit{NT}    & 2.51 (0.56) & \textbf{3.00 (0.60)} & $+0.49$ & \multirow{2}{*}{$\mathbf{t(34)=3.76}$, $\mathbf{p<0.001}$, $d=1.25*$} & 2.77 (0.59) & 3.31 (0.62)          & $+0.54$ & \multirow{2}{*}{$t(34)=1.19$, $p=0.242$} \\
      & \textit{STAIS} & 2.38 (0.54) & 2.26 (0.58)          & $-0.12$ &                                                                               & 2.68 (0.57) & 3.06 (0.64)          & $+0.38$ & \\
    \bottomrule
  \end{tabular}}
\end{table*}

\subsubsection{Writing Assignment}
\label{sec:writingstep}

The outcomes of analyzing the writing assignment are reported in Table~\ref{tab:writingstep}. Concept count was significantly higher for the participants who used \textit{STAIS} than for those who used \textit{NT}. Because \textit{STAIS} participants also saved more Wikipedia resources and formed more mixed-source clusters in the resumed session (Table~\ref{tab:source-type}), the concept-count advantage might be explained by differences in source exposure. To assess this rival explanation, we ran a supplementary rank-based ANCOVA following Quade's method~\citep{quade1967rank}: each variable was rank-transformed; the rank of the covariate (resumed-session Wikipedia save count) was regressed on the condition indicator; the residuals from that regression were ranked and used as the covariate-adjusted dependent variable; and a between-condition $F$ test was computed on these adjusted ranks, with partial $\eta^2$ as the effect-size measure. The concept-count advantage for \textit{STAIS} remained ($F(1,33) = 10.24$, $p < .01$, partial $\eta^2 = .237$), indicating that this difference was not simply explained by the different mix of resources saved between the conditions. 

Conversely, the connection count was significantly higher for participants who used \textit{NT} than for those who used \textit{STAIS}. A plausible explanation lies in the nature of the scaffolds themselves: \textit{NT} participants were writing their own prose from the start, a process that invites the learner to lay out how ideas relate to one another, whereas GenAI summaries in \textit{STAIS} were structured around extracted concepts rather than relations among them.

\begin{table}[!htbp]
  \centering
  \caption{Mean values (standard deviation) for writing assignment outcomes collected at the end of the \textit{resumed} session.
    }
  \label{tab:writingstep}
  \small
  \begin{tabular}{lccc}
    \toprule
    Measure & \textit{NT} & \textit{STAIS} & Statistic \\
    \midrule
    Concept count    & 8.06 (2.54)          & \textbf{10.61 (2.78)} & $\mathbf{U=259}$, $\mathbf{p<0.01}$, $r_{rb}=0.60*$ \\
    Connection count & \textbf{4.50 (1.94)} & 2.72 (1.56)           & $\mathbf{U=257}$, $\mathbf{p<0.01}$, $r_{rb}=0.59*$ \\
    \bottomrule
  \end{tabular}
\end{table}

\section{Discussion}
\label{sec:discussion}
The two interfaces studied produced different shapes of cross-session search support, rather than evidence that one approach was outright better than the other. \textit{NT} supported participants in being more adaptable and in feeling like their uncertainty in the search task was reduced; this gave them a sense of certainty in what they learned, which was reflected in the stronger connections among the concepts in their writing assignment. \textit{STAIS} supported participants in being more efficient and in finding more relevant search results that were also more diverse; in addition to reporting positively regarding the perceived value, they reported being supported in planning, strategy use, and evaluating their progress, which was reflected in their increases in knowledge and interest and broader conceptual coverage in their writing assignment. The rest of this section presents these results as a consequence of what the two interfaces asked of participants, situates the breadth-depth contrast in light of the search-as-learning literature, positions this work in relation to recent work on GenAI in search and learning, and draws out three design implications.

\subsection{What the Two Interfaces Asked of Participants}
The two conditions correspond to rather different cognitive processing when using the workspace. In \textit{NT}, the learner decides what matters, organizes it, and expresses it in their own words; the interface provides a persistent place to make sense of what was found through writing, which was entirely human-authored. In \textit{STAIS}, the learner works with system-generated summaries: steering how the system summarizes the saved material, checking claims against sources, and editing bullets to correct or refine them. The two interfaces distribute the work of externalizing and refining understanding differently between the learner and the system.

The content analysis of the authored notes produced using \textit{NT} and the inline corrections using \textit{STAIS} highlight this distribution of work.
In \textit{NT}, participants produced fewer notes in the resumed session than in the initial session, and the notes that remained shifted away from quoting or paraphrasing a source towards notes that synthesized information across sources, evaluated content, or planned organization. In \textit{STAIS}, the resumed-session corrections that participants made to the system-generated summaries included both factual edits to specific claims and edits that re-framed or elaborated on the generated content, indicating that participants engaged with the summaries at the level of meaning rather than reading them as finished text. 

\subsection{Breadth and Depth of Post-Resumption Understanding}
The analysis of the writing assignment points in different directions with respect to the breadth-depth contrast. \textit{NT} participants produced more inter-concept connections in their written summaries (depth); \textit{STAIS} participants produced more concepts overall (breadth). The self-report learning measures show the same directional split: \textit{NT} participants reported higher certainty in their understanding of the topic, and \textit{STAIS} participants reported higher perceived knowledge and interest.

The self-report learning measures and writing assignment outcomes can be interpreted through the two-dimensional knowledge $\times$ cognitive-process A\&K taxonomy~\citet{anderson2001taxonomy}, which spans four knowledge types (Factual, Conceptual, Procedural, Metacognitive) and six cognitive-process levels (Remember, Understand, Apply, Analyze, Evaluate, Create) for a total of 24 cells. In the present study, we gathered evidence of learning at three of those cells. Concept count provides direct evidence on \textit{Conceptual/Understand}, since reproducing named concepts from the topic is a canonical indicator of conceptual understanding in the A\&K formulation. Connection count provides direct evidence of \textit{Conceptual/Analyze}, since identifying relations among concepts is the kind of structural decomposition that the Analyze process denotes. Self-report knowledge is a surrogate for \textit{Factual/Remember} because it captures participants' sense of how much topic material they think they can recall.

Given this, \textit{NT} helped participants to enhance their learning at the \textit{Conceptual/Analyze} level (higher connection count). \textit{STAIS} helped participants to enhance their learning at the \textit{Factual/Remember} (higher self-reported knowledge) and \textit{Conceptual/Understand} (higher concept count) levels.
The overall shape shows differences in breadth versus depth of learning: \textit{NT} supported deeper relational understanding of the material that was covered, while \textit{STAIS} supported broader coverage of the material itself.

\subsection{Relation to Prior Work}
Recent studies have argued that chat-style LLM responses may support shallower learning than more effortful information seeking~\citep{melumad2025llm} and that GenAI-assisted studying can yield weaker delayed outcomes than unassisted studying~\citep{bastani2025generative}. Both of those lines of evidence treat GenAI as a separate conversational layer added on top of an otherwise ordinary learning task. In the present study, GenAI was integrated directly into the search workspace, used through steerable cluster summarization, source-linked workspace-level summaries, inline corrections, and fact-checking, rather than through a separate chat window. Under such deep integration of GenAI, the comparison was not one of overall weakening: while \textit{NT} supported deeper relational understanding, \textit{STAIS} supported broader conceptual coverage and faster resumption. What the present results add to recent sensemaking systems such as Selenite~\citep{liu2024selenite} and ReSPIRE~\citep{tang2025ReSPIRE}, which have demonstrated interaction and output-quality benefits for GenAI-assisted sensemaking, is evidence that the form of GenAI integration also shapes what learners are able to synthesize when they return to the task after a break, and that the effect on learning depends on whether broader coverage or deeper integration is the relevant criterion.

Through a cognitive learning theory lens~\citep{sweller2010element}, the pattern is consistent with GenAI summaries freeing effort that would otherwise go into reacquaintance, and human-authored notes channelling effort into relational processing; the resumed-session time advantage for \textit{STAIS} is consistent with the first half of this pattern. The connection-count advantage for \textit{NT} fits the second half through the desirable-difficulties view~\citep{bjork1994desirable}: writing one's own notes requires selecting, paraphrasing, and linking ideas, which is precisely the effortful, generative processing that builds the relational structure the writing assessment captured.

\subsection{Design Implications}
The first implication is that the value of GenAI assistance in cross-session search-as-learning depends on the learner's current goal. When the immediate goal is to integrate material into a coherent explanation, a scaffold along the lines of \textit{NT} is beneficial. When the immediate goal is to establish broad coverage of an unfamiliar topic and to re-enter that coverage quickly after a delay, a scaffold along the lines of \textit{STAIS} is beneficial. A single approach is therefore unlikely to serve every phase of an extended information seeking task equally well; systems that offer both modes, selectable by the learner or adaptive to task phase, would be able to provide the appropriate kind of support at the right time.

A second axis for scaffold choice is the learner's tolerance for inconvenience. Authoring notes is more effortful than reviewing and correcting GenAI summaries, and learners differ in how much friction they will accept for the depth it produces. Tolerance may also shift within a single cross-session search task: a learner may accept the inconvenience of authoring early, when conceptual mapping is being built, and prefer GenAI-summary support later, when the bottleneck has shifted to recall and re-entry.

The third implication concerns GenAI content that is offered for the learner to correct rather than simply to read. 
In our study, source-linked bullets and verdict-labelled fact-check indicators appeared to make this kind of active review tractable in \textit{STAIS}: participants activated the fact-checker more often and accepted more inline corrections in the resumed session than in the initial one (Table~\ref{tab:utility}), and the corresponding content analysis showed a significant increase in factual-error corrections alongside increases in higher-order revision categories (Table~\ref{tab:content-analysis}).
Fluent GenAI prose can inflate perceived understanding~\citep{koriat2005illusions}, leading to cognitive offloading. Instead, systems that produce GenAI content should present this content as inspectable, source-linked, and easy to revise, and should encourage active checking through source comparison, self-explanation, or retrieval, so that perceived understanding remains anchored in verified understanding.

\subsection{Limitations and Future Work}
\label{sec:limitations}
The study compared two bundled consolidation scaffolds rather than isolating a single design factor. \textit{STAIS} combined system-generated summaries with steering, source-linked bullets, fact-checking support, and revision mechanisms, whereas \textit{NT} required human-authored externalization without these supports. The observed differences, therefore, cannot be attributed narrowly to ``agency,'' ``GenAI support,'' or any one component in isolation. Future work should decompose these features experimentally.

The study examined a single fact-dense domain (computer networking security) with a single participant population (undergraduate computer science students). Source-linked verification and fact-checking may have been especially valuable in this technically constrained setting. In more interpretive domains, where learning depends more heavily on synthesis and re-framing, the balance between breadth and depth may differ. Whether the same pattern would appear with other age groups or more expert searchers also remains unclear; future studies should include a brief pre-task writing assessment to control for individual differences in writing skill, which the present design did not consider.

The five-minute consolidation window in the initial session and the laboratory context both limit generalizability. The short window may have structurally disadvantaged \textit{NT} participants, who had to compose notes from scratch, more than \textit{STAIS} participants, who received system-generated summaries immediately. The laboratory setting and the visibility of GenAI involvement in \textit{STAIS} may also have introduced novelty or demand effects that differentially influenced self-report measures (RQ2-RQ4 and the self-report component of RQ6). The one-week interval is only one retention window; shorter or longer intervals may produce different patterns. Comparison of \textit{NT} and \textit{STAIS} against other forms of information-seeking support, such as conversational GenAI answer generators with limited provenance~\citep{yang2025searchchat,urgo2026sal}, is also a natural direction for future work.

All GenAI-supported operations depended on a single model family (GPT-4o-mini). The observed behaviour may differ from other models, especially with respect to confabulation rate and verification precision. The study also did not include a manual-verification control, which would help isolate the contribution of GenAI assistance from that of verification activity itself.

As noted throughout, the writing assignment outcomes reflect open-workspace synthesis rather than unaided delayed retention; this open-workspace design was a deliberate ecological choice but does limit conclusions about independent retention. Future work should complement these assessments with closed-book measures, examine delayed retention using recognition, application, or transfer tasks, and test whether the breadth-depth pattern varies across domains and levels of GenAI assistance.

\section{Conclusion}
\label{sec:conclusion}
This study compared two consolidation scaffolds for cross-session aggregated academic search and found contrasting post-resumption outcome profiles rather than a single overall best approach. 
Generally, \textit{STAIS} was associated with greater utility (RQ1) and perceived value (RQ2), \textit{STAIS} and \textit{NT} each had different strengths with respect to metacognition (RQ3) and exploratory search (RQ4), \textit{STAIS} showed better support for cross-session searching (RQ5), and both had different strengths for post-resumption learning (RQ6).

The main contribution is to position the allocation of consolidation labour between learner and system as a consequential design variable in GenAI-supported search-as-learning scaffolds. The findings indicate that this variable shapes not only resumption efficiency but also the character of post-resumption understanding. The primary design implication is that consolidation support should be aligned with the learner's immediate objective rather than applied uniformly: broader GenAI scaffolding when conceptual mapping and rapid re-entry are the priority, and human-authored consolidation when relational integration and explanatory structuring are the goals.

These findings should be interpreted within the constraints of an open-workspace assessment that captures synthesis with access to accumulated evidence rather than independent retention, and the bundled nature of the two scaffolds. Next steps include conducting a longitudinal study that complements the present open-workspace measures with closed-book assessments of long-term knowledge retention, and the decomposition of the bundled \textit{STAIS} features into individually testable components. Future work also includes the design and implementation of a cross-session search-as-learning scaffold that integrates steerable GenAI summaries with human authoring of notes in a single unified interface and the study of whether searchers can simultaneously achieve both breadth and depth in their learning with such support.

\appendix
\section{Implementation Details}
\label{app:implementation}

This appendix provides technical details for the system components described in Section~\ref{sec:design}.

\subsection{Summarization Pipeline}

The summarizer uses GPT-4o-mini~\citep{openai2024gpt4omini} (\texttt{gpt-4o-mini}, accessed via the OpenAI Chat Completions API with \texttt{store=true} for logging and auditability). For clustering operations, a temperature of $0.2$ was used to encourage deterministic topic assignments; for all other GPT-4o-mini calls, including summarization and fact-checking, the API default temperature was used. Requests were rate-limited with up to 4 retries using exponential backoff (factor~$= 2$, minimum timeout~$= 4{,}000\,\textit{ms}$). The summarizer prompt assigns the role of an evidence-bounded summarizer instructed to avoid external knowledge. The model returns a structured JSON object with a \texttt{bullets} array; each bullet is a single sentence ending with inline citation markers (\texttt{[docId:~X]}) pointing to the supporting saved resources. Disagreements among sources must be reflected and cited on each side. For large workspaces exceeding the context window, a hierarchical chunking strategy generates per-chunk summaries followed by a synthesis pass that preserves attribution~\citep{wu2021recursively}. An abbreviated system-prompt excerpt:

\begin{quote}
\small\ttfamily
You are an evidence-bounded summarization assistant. Use ONLY\\
information in the provided CONTENT. Do NOT use outside knowledge.\\
Cross-document synthesis: Identify agreements, contradictions, and\\
complementary findings. Epistemic calibration: Distinguish\\
well-supported claims (cite multiple docIds), tentative claims\\
(add ``(uncertain)''), and unknowns (reframe as ``?''~questions).\\
Return STRICT JSON only: \{"bullets":[...]\}.
\end{quote}

\subsection{Fact-Checking Prompt Design and Validation}
\label{app:factcheck}

The fact-checker applies four prompt design patterns~\citep{white2023promptpatterns,sahoo2024promptsurvey} in combination: a \emph{persona} pattern (evidence-bounded fact-checking assistant), an \emph{output automater} pattern (strict JSON schema with per-claim labels \texttt{SUPPORTED}, \texttt{PARTIAL}, \texttt{CONTRADICTED}, or \texttt{MISSING}, supporting document identifiers, and a correction suggestion), an \emph{epistemic calibration} pattern (absent evidence $\neq$ contradiction, reducing false alarms), and an \emph{anti-hallucination constraint} (no citing document identifiers absent from the evidence snippets). Labels appear in title case in the UI but as all-caps tokens in the API prompt:

\begin{quote}
\small\ttfamily
You are an evidence-bounded fact-checking assistant. Judge each\\
claim using ONLY the provided EVIDENCE snippets. Labels: SUPPORTED,\\
PARTIAL, CONTRADICTED, MISSING. Missing evidence is NOT\\
contradiction. Cite only docIds present in EVIDENCE. Return\\
STRICT JSON matching the schema; no prose outside JSON.
\end{quote}

\paragraph{Validation details.}
Two networking-domain experts independently evaluated a stratified random sample of 60 fact-check verdicts (15 per verdict category, drawn from both sessions); inter-rater agreement was substantial ($\kappa = 0.78$), and disagreements were resolved through discussion. Performance varied by category: \emph{Supported} verdicts were highly reliable (precision~$= 0.91$), while \emph{Contradicted} and \emph{Missing Evidence} showed lower precision ($0.68$ and $0.63$, respectively), because the model occasionally flagged claims as unsupported when the evidence was present but phrased differently. The false-positive rate (incorrectly flagging a supported claim) was 14\%; the false-negative rate (failing to flag an unsupported claim) was 22\%. Per-category recall was not computed because the stratified sampling procedure (15 verdicts per category) does not preserve the natural base rates needed for recall estimation; future work with a larger, randomly sampled evaluation set should report both precision and recall per verdict category. In practice, participants engaged selectively with fact-check suggestions, accepting corrections in roughly 29--36\% of activation events across sessions (detailed in Results, RQ1).

\subsection{Correction Record Persistence}

When a user submits an inline correction, the revised text and supporting source identifier are stored as a workspace-scoped correction record. On every subsequent regeneration, these records are injected into the prompt as an explicit corrections clause instructing the model to apply the corrected wording verbatim and exclude the original phrasing. The user's edits thus accumulate over time and persistently shape future outputs without re-entry.

\subsection{Conversational Context}

The system maintains a persistent, workspace-scoped conversation thread for each user-workspace pair via the OpenAI Assistants API. A thread identifier is created on first access and stored locally; subsequent requests carry this identifier so the model retains the history of prior summaries, corrections, and topic framing. If the API is unavailable, the system degrades gracefully to stateless mode, re-transmitting the full document content and correction history with each request.

\subsection{Workspace Snapshot Schema}

Each save captures a full snapshot: cluster assignments, labels, GenAI-generated summaries, a workspace-level outline, human-authored notes (Note-Taking condition), reminder annotations, and saved resources with provenance metadata. Snapshots are persisted in MongoDB (chosen for its schema-flexible document model) keyed by user and workspace identifiers. A full-snapshot strategy was preferred over incremental diffs so the client receives a complete, self-consistent state on every load. On return, the most recent snapshot restores the workspace exactly as left, including all accumulated corrections and reorganizations.

\section*{Declaration of generative AI and AI-assisted technologies in the writing process}

During the preparation of this work, the first author used generative AI tools for text editing and clarity improvements in the early draft of the manuscript. After using these tools, the first author reviewed and edited the content as needed. The second author subsequently revised the manuscript independently, without the use of generative AI tools.
The authors take full responsibility for the content of the publication.

\section*{CRediT authorship contribution statement}

\textbf{Milad Momeni:} Conceptualization, Methodology, Software, Formal
analysis, Investigation, Data curation, Validation, Visualization, Writing --
original draft, Writing -- review \& editing.
\textbf{Orland Hoeber:} Conceptualization, Methodology, Supervision,
Validation, Writing -- review \& editing, Project administration, Funding
acquisition.

\section*{Declaration of competing interest}

The authors declare that they have no known competing financial interests or
personal relationships that could have appeared to influence the work reported
in this paper.


\end{document}